\documentclass[runningheads]{llncs}
\usepackage{graphicx}
\usepackage{color}
\usepackage{okcategory}
\usepackage{mathpartir}
\usepackage{tocvsec2}  

\newcommand\gdefinedby{::=}

\newcommand{\KI}{\mathfrak{L} }

\newcommand{\defeqq}{\stackrel {\mathrm{def}}=}
\newcommand{\defeq}{\stackrel {\mathrm{def}}=}

\newcommand{\initiall}{\mathsf{0}}
\newcommand{\epzeroo}{\mathfrak{O}}

\newcommand{\uniqMor}{\iota}
\newcommand{\terminall}{\mathsf{1}}

\newcommand{\wcpo}{$\omega$-cpo}

\newcommand{\wCpo}{\mathbf{\boldsymbol\omega Cpo}}
\renewcommand{\wCPO}{\mathbf{\boldsymbol\omega CPO}}

\newcommand\epto{\stackrel{\hookrightarrow}{\leftharpoondown}}

\newcommand\eppair[1]{\left( \emb #1 , \prj #1 \right) }
\newcommand\emb[1]{#1^e}
\newcommand\prj[1]{#1^p}
\newcommand\Positive[2]{\mathfrak{P}_{#2}^{\mathfrak{d}0}\left( #1 \right)}

\newcommand\lift[1]{\overline{#1}}
\newcommand\unlift[1]{\underline{#1}}

\newcommand\unlifteppair[1]{\hat{{#1}}}
\newcommand\secondM[1]{#1 ^\ast }

\newcommand\semanticshandler[2]{\mathfrak{p}_{#1 \to #2} }

\newcommand\ifelse[3]{\mathbf{if}\,#1\,\mathbf{then}\,#2\,\mathbf{else}\,#3\,}
\newcommand\letin[3]{\mathbf{let}\,#1=\,#2\,\mathbf{in}\,#3}

\newcommand\intle[1]{\varphi_{#1}}

\newcommand\coproje[1]{\iota _{#1} }

\newcommand\cpairL{\langle }
\newcommand\cpairR{\rangle }

\newcommand\TyAlph[1]{
	\ifcase #1\or \tau\or \sigma\or \rho\else \@ctrerr \fi%
}
\newcommand\ty[1]{{\TyAlph{#1}}}

\newcommand\TySchAlph[1]{
	\ifcase #1\or \phi\or \psi \or \upsilon\else \@ctrerr \fi%
}

\newcommand\var[1]{{\VarAlph{#1}}}
\newcommand\VarAlph[1]{%
	\ifcase #1\or x\or y\or z\else \@ctrerr \fi%
}

\newcommand\tvar[1]{{\TVarAlph{#1}}}
\newcommand\TVarAlph[1]{%
	\ifcase #1\or \alpha\or \beta\or \gamma\else \@ctrerr \fi%
}

\newcommand\val[1]{%
	\ifcase #1\or v\or w\or u\else \@ctrerr \fi%
}

\newcommand\trm[1]{{\TermAlph{#1}}}
\newcommand\TermAlph[1]{%
	\ifcase #1\or t\or s\or r\else \@ctrerr \fi%
}

\newcommand\monadLR[2]{\mathcal{P}_{#1}\monadwP{#2} }
\newcommand\openLift[1]{\mathsf{Diff}_{\left( #1 \right) } } 
\newcommand\topologyO[1]{\mathfrak{O}_{#1}}

\newcommand\morLRmonad[1]{\mathtt{j}_{#1}}

\newcommand\wrapSyncat[1]{\mathbf{wrap}_{#1}}

\newcommand\ee{\eta } 
\newcommand\mm{\mathrm{m}} 
\newcommand\ID{\mathrm{id}} 

\newcommand\RRsyntax{\mathtt{R}}
\newcommand\monadT{\mathcal{T}} 

\newcommand\monadS{\mathcal{S}} 
\newcommand\monadwP[1]{\left( #1\right) _\leastelement } 

\newcommand\catV{\Cat{V}}

\newcommand\catB{\Cat{B}}
\newcommand\catC{\Cat{C}}
\newcommand\catD{\Cat{D}}

\newcommand\RCBVcat{\mathfrak{C}_{\mathcal{RBV}} }

\newcommand\wrCBVcat{\wCPO\textrm{-}\mathfrak{C}_{r\mathcal{BV}} }

\newcommand\SUBscone[2]{\mathbf{Sub}\left( #1\downarrow #2 \right)}

\newcommand\obb[1]{\mathsf{ob}\, #1}

\newcommand\canonicalbasise[2]{e^{#1}_{#2}}
\newcommand\sconeFUNCTOR[1]{G_{#1}}

\newcommand\CBVU{\mathcal{U}_{\mathcal{BV}} }
\newcommand\rCBVU{\mathcal{U}_{r\mathcal{BV}} }

\newcommand\forgetfulS{\mathcal{L}}

\newcommand\forgetfulSub{\underline{\mathcal{L}}}

\newcommand\ihom[3]{#1\left[ #2 , #3 \right] }

\newcommand\ihomV{\ihom{\catV} }

\newcommand\ehom[3]{#1\left( #2 , #3 \right) }

\newcommand\op{\mathrm{op} }  

\newcommand\rolling{\unlift{\rollReT} }

\newcommand\pE[2]{{#1}_{#2}}
\newcommand\mind[1]{\treeindexing{m}{#1}}

\newcommand\treeindexing[2]{{\mathsf{#1}_{\left( #2\right) }}}

\newcommand\underDot[1]{\underline{\mathtt{#1}}}
\newcommand\coloneq{\colon =}
\newcommand\pEE[1]{#1}

\newcommand\ParamVdtype[2]{\mathfrak{P}_{#2}^{\mathfrak{d}}\left( #1 \right)  }

\newcommand\ParamAll[2]{\mathsf{Param}\left( #1 , #2\right)  }

\newcommand\rollReT{\mathsf{roll} } 
 
\newcommand\pwrollReT[1]{\omega\mathsf{roll}^{#1}  } 

\newcommand\ffixpointRecT{\unlift{\fixpointRecT} } 
\newcommand\wffixpointRecT{\unlift{\fixpointRecT}_{\omega } }

\newcommand\fixpointRecT{\nu }
\newcommand\wfixpointRecT{\nu _{\omega} }

\newcommand\leastelement{\bot}                              

\newcommand\diagk[1]{\mathrm{diag}_{#1} }
\newcommand\proj[1]{\pi_{#1}}            

\newcommand\pairL{\left( }
\newcommand\pairR{\right) }

\newcommand{\RR}{\mathbb{R}}
\newcommand{\NN}{\mathbb{N}}
\newcommand\NNN[1]{\mathbb{I}_{#1}}

\newcommand{\signR}{\mathsf{sign}}
\newcommand\semanc[1]{\mathsf{#1} }
\newcommand\seman[1]{ {#1}_\op }

\newcommand\targetreals{\reals }  
\newcommand\tangentprojection[2]{\mathfrak{h} _{#1} #2 }
\newcommand\tangentreals{\mathbf{vect} }  
\newcommand\ints{\mathbf{int}}     
\renewcommand\reals{\mathbf{real}}            
\newcommand\cnst[1]{\underline{#1}}           
\newcommand\cnzero{\overline{0}}
\newcommand\cncanoni[1]{\overline{e} _{#1}}

\newcommand\Op{\mathrm{Op}}                   

\newcommand\sigmoid{\varsigma}

\newcommand\syncat[1]{\mspace{-25mu}\synname{#1}}
\newcommand\synname[1]{\qquad\text{#1}}

\newenvironment{syntax}[1][]{%
	\(
	\begin{array}[t]{#1l@{\quad\!\!}*3{l@{}}@{\,}l}
	}{
	\end{array}
	\)%
}

\newcommand\gor{\mathrel{\lvert}}
\newcommand\vor{\mathrel{\big\lvert}}

\newcommand\Init{\mathbf{0}}

\newcommand\tTuple[1]{\langle #1\rangle}
\newcommand\tUnit{\tTuple{\,}}

\newcommand\pMatch[5][\,]{\mathbf{case}\,#2\,\mathbf{of}#1\tPair{#3}{#4}\To#5}

\newcommand\vMatch[3][\,]{\mathbf{case}\,#2\,\mathbf{of}#1\{#3\}}
\newcommand\nvMatch[2][\,]{\mathbf{case}\,#2\,\mathbf{of}#1\{\,\}}
\newcommand\bvMatch[6][\,]{\mathbf{case}\,#2\,\mathbf{of}#1\{\tInl#3\To #4\, \mid \tInr#5\To #6\}}
\newcommand\rMatch[4][\,]{\mathbf{case}\,#2\,\mathbf{of}#1\mathbf{roll}\,#3\To#4}
\newcommand\tItFrom[3]{\mathbf{iterate}\,#1\,\mathbf{from}\,#2=#3}
\newcommand\To{\to}
\newcommand\tRoll{\mathbf{roll}\,}

\newcommand\tInl{\mathbf{inl}\,}
\newcommand\tInr{\mathbf{inr}\,}
\newcommand\tFst{\mathbf{fst}\,}

\newcommand\tPair[2]{\langle #1, #2\rangle}

\newcommand\trec[2]{\mathbf{\mu}#1.#2}

\WithSuffix\newcommand\t*{\,{\mathop{\times}}\,}
\WithSuffix\newcommand\t+{\,{\mathop{\sqcup}}\,}
\newcommand\tSign{\mathbf{sign}\,}
\newcommand\Unit{\mathbf{1}}
\newcommand\fun[1]{\lambda #1.}

\newcommand\subst[2]{#1{}[#2]}
\newcommand\sfor[2]{^{#2}\!/\!_{#1}}

\newcommand\tinf{\vdash}
\newcommand\ctx{\Gamma}
\newcommand\Ginf[3][]{\ctx #1\tinf #2 : #3}

\newcommand\DGinf[3][]{\Delta\mid\ctx #1\tinf #2 : #3}

\newcommand\freeeq[1]{\stackrel{\# #1}{=}}

\newcommand\Dsynsymbol[1][]{\scalebox{0.8}{$\mathcal{D}$}_{#1}}
\newcommand\Dsyn[2][]{\Dsynsymbol[#1](#2)}

\newcommand\vectoraslineartransformation[1]{\tilde{#1}}

\newcommand\Dsynplainsymbol[1][]{\scalebox{0.8}{${\mathcal{D}}$}_{#1}}
\newcommand\Dsynplain[2][]{\Dsynplainsymbol[#1](#2)}

\newcommand\DSyn{\DSynrec}
\newcommand\DSynrec{\mathbb{D}}

\newcommand\dDSyn[1]{\overrightarrow{d}#1 }
\newcommand\SEMLRR[2]{\sem{#1}_{#2}}
\newcommand\dDSemtotaltra[2]{\mathfrak{D}^{#1}{#2} }

\newcommand\dDSemtra[2]{\mathfrak{d}^{#1}\left( #2\right)  }
\newcommand\dDSemj[2]{\mathfrak{d}_{#2}\left( #1\right)  }

\newcommand\SynVt{\Syn_V^{\target}}

\newcommand\SynTt{\Syn_\monadS^{\target}}

\newcommand\target{{\mathbf{tr}}}

\newcommand\SynVr{\SynV}

\newcommand\SynTr{\SynT}

\newcommand\SynffixpointRecT{\ffixpointRecT_\Syn}

\newcommand\SynVrt{\SynVt}

\newcommand\SynTrt{\SynTt}
\newcommand\SyntffixpointRecT{\ffixpointRecT^{\target}_\Syn}

\newcommand\PTYPE{{\mathsf{T}_p}}
\newcommand\LRMONAD[1]{\monadT_{#1}}
\newcommand\Syn{\mathbf{Syn}}
\newcommand\SynV{\Syn_V}

\newcommand\SynT{\Syn_\monadS}

\newcommand\semt[2]{[\hspace{-1.5pt}[#2]\hspace{-1.5pt}] _ {#1} }
\newcommand\sem[1]{[\hspace{-1.5pt}[#1]\hspace{-1.5pt}] }
\newcommand\semLR[2]{ \overline{[\hspace{-1.5pt}\sem{#2}\hspace{-1.5pt}]}_{#1} }

\newcommand\extendedH[1]{\mathcal{#1}}

\makeatletter
\let\c@equation\c@figure    
\makeatother
\usepackage[shortlabels]{enumitem}
\makeatletter
\@addtoreset{equation}{section}
\makeatother
\numberwithin{equation}{section}
\counterwithin{figure}{section}

\usepackage{ifluatex}
\input diagxy        
\xyoption{curve}     
\xyoption{all}
\xyoption{2cell}
\allowdisplaybreaks
\begin{document}
\title{Logical Relations for Partial Features and\\ Automatic Differentiation Correctness
	\thanks{This project has received funding via NWO Veni grant number VI.Veni.202.124
		as well as the European Union’s Horizon 2020 research and innovation
		programme under the Marie Skłodowska-Curie grant agreement No. 895827. 
				This research was supported through the programme ``Oberwolfach Leibniz Fellows'' by the Mathematisches Forschungsinstitut Oberwolfach in 2022. It was also partially supported  by the CMUC, Centre for Mathematics of the University of Coimbra - UIDB/00324/2020, funded by the Portuguese Government through FCT/MCTES. }
	}
	
\titlerunning{Logical Relations for Partial Features and AD Correctness}
	
\author{Fernando Lucatelli Nunes\orcidID{0000-0002-1817-2797} \and\\
		Matthijs V\'ak\'ar\orcidID{0000-0003-4603-0523}}
	\authorrunning{F. Lucatelli Nunes \and M.I.L. V\'ak\'ar}
	%
		\institute{Utrecht University\\
		\email{\{f.lucatellinunes,m.i.l.vakar\}@uu.nl}}
	\maketitle              
	\begin{abstract}
		
		We present a simple technique for semantic, open logical relations arguments about languages with recursive types, which, as we show, follows from a principled foundation in categorical semantics.
		We demonstrate how it can be used to give a very straightforward proof of correctness of practical forward- and reverse-mode dual numbers style automatic differentiation (AD) on ML-family languages.
		The key idea is to combine it with a suitable open logical relations technique for reasoning about differentiable partial functions (a suitable lifting of the partiality monad to logical relations), which we introduce.
		
	\end{abstract}

	
	\ifluatex
	\directlua{adddednatlualoader = function ()
	require = function (stem)
	local fname = dednat6dir..stem..".lua"
	package.loaded[stem] = package.loaded[stem] or dofile(fname) or fname
	end
	end}
	\catcode`\^^J=10
	\directlua{dofile "dednat6load.lua"}
	\else
	%
	\def\diagxyto{\ifnextchar/{\toop}{\toop/>/}}
	\def\to     {\rightarrow}
	\def\defded#1#2{\expandafter\def\csname ded-#1\endcsname{#2}}
	\def\ifdedundefined#1{\expandafter\ifx\csname ded-#1\endcsname\relax}
	\def\ded#1{\ifdedundefined{#1}
		\errmessage{UNDEFINED DEDUCTION: #1}
		\else
		\csname ded-#1\endcsname
		\fi
	}
	\def\defdiag#1#2{\expandafter\def\csname diag-#1\endcsname{\bfig#2\efig}}
	\def\defdiagprep#1#2#3{\expandafter\def\csname diag-#1\endcsname{{#2\bfig#3\efig}}}
	\def\ifdiagundefined#1{\expandafter\ifx\csname diag-#1\endcsname\relax}
	\def\diag#1{\ifdiagundefined{#1}
		\errmessage{UNDEFINED DIAGRAM: #1}
		\else
		\csname diag-#1\endcsname
		\fi
	}
	\newlength{\celllower}
	\newlength{\lcelllower}
	\def\cellfont{}
	\def\lcellfont{}
	\def\cell #1{\lower\celllower\hbox to 0pt{\hss\cellfont${#1}$\hss}}
	\def\lcell#1{\lower\celllower\hbox to 0pt   {\lcellfont${#1}$\hss}}
	\def\expr#1{\directlua{output(tostring(#1))}}
	\def\eval#1{\directlua{#1}}
	\def\pu{\directlua{pu()}}
	%
	
\defdiag{obvious-diagram-parametric-types-n}{   
	\morphism(0,-300)|l|/->/<0,300>[{{\left({\catV}^\op\times\catV\right)^{n+1}}}`{{\left({\catC}^\op\times\catC\right)^{n+1}}};{{\left({J^\op}\times{J}\right)^{n+1}}}]
	\morphism(675,-300)|a|/->/<0,300>[{{\catV}}`{{\catC}};{{J}}]
	\morphism(0,-300)|b|/->/<675,0>[{{\left({\catV}^\op\times\catV\right)^{n+1}}}`{{\catV}};{{\pE{E}{\catV}}}]
	\morphism(0,0)|a|/->/<675,0>[{{\left({\catC}^\op\times\catC\right)^{n+1}}}`{{\catC}};{{\pE{E}{\catC}}}]
}

\defdiag{roll-basic-diagram}{   
	\morphism(0,0)|a|/->/<1125,0>[{{\left({\catV}^\op\times\catV\right)^{n}}}`{{\left({\catV}^\op\times\catV\right)^{n+1}}};{{\pairL{\id},\pE{\fixpointRecT{E}}{\catV}^\op{,}\pE{\fixpointRecT{E}}{\catV}\pairR}}]
	\morphism(0,0)|b|/->/<1125,-450>[{{\left({\catV}^\op\times\catV\right)^{n}}}`{{\catV}};{{\pE{\fixpointRecT{E}}{\catV}}}]
	\morphism(1125,0)|r|/->/<0,-450>[{{\left({\catV}^\op\times\catV\right)^{n+1}}}`{{\catV}};{{\pE{E}{\catV}}}]
	\morphism(1125,-450)|b|/->/<-1125,0>[{{\catV}}`{{\catC}};{{J}}]
	\morphism(675,-225)|a|/<=/<375,0>[{\phantom{O}}`{\phantom{O}};{{\rollReT}^{\pEE{E}}}]
}

\defdiag{obvious-diagram-parametric-types-(n-1)-recursive}{   
	\morphism(0,-300)|l|/->/<0,300>[{{\left({\catV}^\op\times\catV\right)^{n}}}`{{\left({\catC}^\op\times\catC\right)^{n}}};{{\left({J^\op}\times{J}\right)^{n}}}]
	\morphism(675,-300)|a|/->/<0,300>[{{\catV}}`{{\catC}};{{J}}]
	\morphism(0,-300)|b|/->/<675,0>[{{\left({\catV}^\op\times\catV\right)^{n}}}`{{\catV}};{{\pE{\fixpointRecT{E}}{\catV}}}]
	\morphism(0,0)|a|/->/<675,0>[{{\left({\catC}^\op\times\catC\right)^{n}}}`{{\catC}};{{\pE{\fixpointRecT{{E}}}{\catC}}}]
}

\defdiag{roll-eq1-diagram}{  
	\morphism(0,0)|a|/->/<1050,0>[{{\left({\catV}^\op\times\catV\right)^{n-1}}}`{{\left({\catV}^\op\times\catV\right)^n}};{{\pairL{\id},\pE{\fixpointRecT{E}}{\catV}^\op{,}\pE{\fixpointRecT{E}}{\catV}\pairR}}]
	\morphism(0,0)|b|/->/<1050,-450>[{{\left({\catV}^\op\times\catV\right)^{n-1}}}`{{\catV}};{{\pE{\fixpointRecT{E}}{\catV}}}]
	\morphism(1050,0)|r|/->/<0,-450>[{{\left({\catV}^\op\times\catV\right)^n}}`{{\catV}};{{\pE{E}{\catV}}}]
	\morphism(1050,-450)|b|/->/<-1050,0>[{{\catV}}`{{{\catV}'}};{{H}}]
	\morphism(638,-225)|a|/<=/<375,0>[{\phantom{O}}`{\phantom{O}};{{\rollReT}^{\pEE{E}}}]
}

\defdiag{roll-eq2-diagram}{   
	\morphism(0,0)|a|/->/<1125,0>[{{\left({{\catV}'}^\op\times{\catV}'\right)^{n-1}}}`{{\left({{\catV}'}^\op\times{\catV}'\right)^n}};{{\pairL{\id},\pE{\fixpointRecT{E'}}{{\catV}'}^\op{,}\pE{\fixpointRecT{E'}}{{\catV}'}\pairR}}]
	\morphism(0,0)|b|/->/<1125,-450>[{{\left({{\catV}'}^\op\times{\catV}'\right)^{n-1}}}`{{{\catV}'}};{{\pE{\fixpointRecT{E'}}{{\catV}'}}}]
	\morphism(1125,0)|r|/->/<0,-450>[{{\left({{\catV}'}^\op\times{\catV}'\right)^n}}`{{{\catV}'}};{{\pE{E}{{\catV}'}}}]
	\morphism(0,-450)|l|/->/<0,450>[{{\left({\catV}^\op\times\catV\right)^{n-1}}}`{{\left({{\catV}'}^\op\times{\catV}'\right)^{n-1}}};{{\left({H}^\op\times{H}\right)^{n-1}}}]
	\morphism(675,-225)|a|/<=/<375,0>[{\phantom{O}}`{\phantom{O}};{{\rollReT}^{\pEE{E'}}}]
}

\defdiag{diag-H-compatibility-of-paramatric-types}{  
	\morphism(0,0)|a|/->/<725,0>[{{\left({\catV}^\op\times\catV\right)^{n+1}}}`{{\catV}};{{\pE{E}{{\catV}}}}]
	\morphism(0,0)|b|/->/<0,-300>[{{\left({\catV}^\op\times\catV\right)^{n+1}}}`{{\left({\catV'}^\op\times\catV'\right)^{n+1}}};{{\left(H^\op\times{H}\right)^{n+1}}}]
	\morphism(725,0)|r|/->/<0,-300>[{{\catV}}`{{{\catV}'}};{{H}}]
	\morphism(0,-300)|b|/->/<725,0>[{{\left({\catV'}^\op\times\catV'\right)^{n+1}}}`{{{\catV}'}};{{\pE{E'}{{\catV}'}}}]
}

\defdiag{morphism-comma-category}{  
	\morphism(0,0)|l|/->/<0,-225>[{D}`{{G_{\ty{1}}(C)}};{{j}}]
	\morphism(0,0)|a|/->/<900,0>[{D}`{{D}'};{{{\alpha}_0}}]
	\morphism(900,0)|r|/->/<0,-225>[{{D}'}`{{G_{\ty{1}}(C')}};{{h}}]
	\morphism(0,-225)|b|/->/<900,0>[{{G_{\ty{1}}(C)}}`{{G_{\ty{1}}(C')}};{{G_{\ty{1}}\left(\alpha_1\right)}}]
}

\defdiag{basic-logicalrelations-recursive-types}{   
	\morphism(0,0)|a|/->/<1750,0>[{{\left(\SynVr,\SynTr{,}\SynffixpointRecT\right)}}`{{{\left(\SynVr,\SynTr{,}\SynffixpointRecT\right)}\times\left(\SynVrt{,}\SynTrt{,}\SyntffixpointRecT\right)}};{{\left(\ID{,}\DSynrec\right)}}]
	\morphism(1750,0)|r|/->/<0,-300>[{{{\left(\SynVr,\SynTr{,}\SynffixpointRecT\right)}\times\left(\SynVrt{,}\SynTrt{,}\SyntffixpointRecT\right)}}`{{\rCBVU\left(\wCpo\times\wCpo{,}\monadwP{-}\right)}};{{\sem{-}\times\semt{k}{-}}}]
	\morphism(0,0)|l|/->/<0,-300>[{{\left(\SynVr,\SynTr{,}\SynffixpointRecT\right)}}`{{\rCBVU\left(\SUBscone{\wCpo}{\sconeFUNCTOR{n}},\monadLR{n}{-}\right)}};{{\semLR{n}{-}}}]
	\morphism(0,-300)|b|/->/<1750,0>[{{\rCBVU\left(\SUBscone{\wCpo}{\sconeFUNCTOR{n}},\monadLR{n}{-}\right)}}`{{\rCBVU\left(\wCpo\times\wCpo{,}\monadwP{-}\right)}};{{\rCBVU\left({\forgetfulSub}_{n}\right)}}]
}
\section{Introduction}\label{sec:introduction}
Automatic differentiation (AD) computes derivatives in a numerically stable way that scales efficiently to high-dimensional spaces. Its ubiquity in scientific computing, statistics and machine learning applications has led to the idea of differentiable programming: compilers for modern programming languages should provide good built-in support for AD of any program written in those languages \cite{meijer2018behind,plotkin2018some}.
It is a non-trivial question how to efficiently and correctly differentiate arbitrary complex programs, which has led to a booming area of research.

Dual numbers techniques give very simple forward and reverse mode AD algorithms for expressive ML-family functional languages \cite{shaikhha2019efficient,brunel2019backpropagation,DBLP:conf/fossacs/HuotSV20,DBLP:journals/pacmpl/MazzaP21,DBLP:journals/pacmpl/KrawiecJKEEF22,smeding2022}.
The correctness proofs for these algorithms  rely on (open) semantic logical relations arguments.
These proofs become surprisingly complex, even for simple dual numbers AD algorithms, when considering AD of languages with partial features such as recursion.
For example, the authors have given such proofs in the past (in the preprint \cite{vakar2020denotational}), but despite our best efforts they seemed to require subtle combinations of sheaves (to deal with the surprisingly hard interaction between partiality and differentiation induced by conditionals on real numbers) and \wcpo-structure (required for recursion). 
The resulting proofs were dissatisfying due to their complexity, restricting the audience and obstructing their generalisation to more advanced AD algorithms like CHAD  \cite{DBLP:journals/toplas/VakarS22,VAKAR-LUCATELLI2021}.

The present work is a vast simplification of the arguments in \cite{vakar2020denotational}
(and, also, \cite{DBLP:conf/fossacs/HuotSV20}), not requiring any (\wcpo{s} internal to) sheaves or diffeological spaces and instead relying on plain \wcpo{s}.
This simplification is desirable if we are to apply the techniques more generally, such as to the correctness of CHAD \cite{DBLP:journals/toplas/VakarS22,VAKAR-LUCATELLI2021} for recursion,
and if we want the technique to be more broadly accessible.

The first contribution of this paper lies in the development of very simple but powerful, semantic logical relations techniques for reasoning about recursive types.
By contrast with the existing techniques of \cite{pitts1996relational,DBLP:conf/esop/Ahmed06}, our technique follows directly from standard category theoretic recipes and applies equally to open logical relations (sconing along functors $G$ other than $\Hom{1,-}$)~\cite{bcdg-open-logical-relations}.

The second contribution is the development of a simple logical relations technique for reasoning about partially defined differentiable functions, which can be understood as a particular lifting of the partiality monad to our logical relations.

Thirdly, we show that the combination of these two techniques suffices to give a very elegant correctness proof of the practically useful dual-numbers style AD algorithms of expressive ML-family languages implemented by \cite{smeding2022}.

Finally, our arguments can be generalised to the correctness argument of the more advanced AD technique CHAD when applied to languages with partial features, which motivates a lot of this development.

\section{Why we care about differentiating partial programs?}\label{sec:why-differentiate-expressive}
Given the central role that AD plays in modern scientific computing and
machine learning, the ideal of differential programming is emerging \cite{meijer2018behind,plotkin2018some}:
compilers for general purpose programming languages should provide built-in support for automatic differentiation of any programs written in the language.
What a correct and efficient notion of derivative is of some popular programming language features might not be so straightforward, however, as they often go beyond what is studied in traditional calculus.
In this paper we focus on the challenge posed, in particular, by partial language features: partial primitive operations, lazy conditionals on real numbers, iteration, recursion and recursive types.

Partial primitive operations are key: even the basic operations of division and logarithm are examples.
(Lazy) conditionals on real numbers are used to paste together existing smooth functions, as basic example being the ReLU function\vspace{-6pt}
$$
ReLU(x) \defeq \ifelse{x}{0}{x}=\vMatch{(\tSign\,x)}{\tInl\_\To 0\mid \tInr\_\To x},\vspace{-4pt}
$$
which is a key component of many neural networks.
They are also frequently used in probabilistic programming to paste together density functions of different distributions \cite{betancourt_2019}.
People have long studied the subtle issue of how one should algorithmically differentiate such functions with ``kinks'' under the name of \emph{the if-problem in automatic differentiation} \cite{beck1994if}.  
Our solution is the one also employed  by \cite{abadi-plotkin2020}: to treat the functions as semantically undefined at their kinks (at $x=0$ in the case of $ReLU(x)$).
This is justified given how coarse the semantic treatment of floating point numbers as real numbers is already.
Our semantics based on partial functions defined on real numbers is sufficient to prove many high-level correctness properties. 
However, like any semantics based on real numbers, it fails to capture many of the low-level subtleties introduced by the floating point implementation.
Our key insight that we use to prove correctness of AD of partial programs is to construct a suitable lifting of the partiality monad to a variant of \cite{DBLP:conf/fossacs/HuotSV20}'s category of $\RR^k$-indexed logical relations used to relate programs to their derivatives.
This particular monad lifting for derivatives of partial functions can be seen as our solution to the if-problem in AD.

Similarly, iteration constructs, or while-loops, are necessary for implementing iterative algorithms with dynamic stopping criteria.
Such algorithms are frequently used in programs that AD is applied to.
For example, AD is applied to iterative differential equation solvers to perform Bayesian inference in
SIR models.
This technique played a key role in modelling the Covid19-pandemic \cite{flaxman2020estimating}.
For similar reasons, AD through iterative differential equation solvers 
is important for probabilistic modelling of  pharmacokinetics \cite{tsiros2019population}.
Other common use-cases of iterative algorithms that need to be
AD'ed are eigen-decompositions and algebraic equation solvers, such as those employed in Stan \cite{carpenter2015stan}. 
Finally, iteration gives a convenient way of achieving numerically stable approximations to complex functions (such as the  Conway-Maxwell-Poisson density function \cite{goodrich_2017}).
The idea is to construct, using iteration, a Taylor approximation that terminates once the next term in the series causes floating-point underflow.
Indeed, for a function whose $i$-th terms in the Taylor expansion can be represented by a program\vspace{-8pt}
$$i : \ints, x : \reals \vdash
t(i, x) : \reals,\vspace{-2pt}$$
we would define the underflow-truncated Taylor series by
$$\tItFrom{\Bigg(\begin{array}{l}
\pMatch{ x}{x_1}{x_2}{}
\letin{y}{
t(x_1, x_2)}{}\\
\vMatch{-c < y < c}{\tInl\_ \To \tInr x_2\\
\hspace{88pt}\mid \tInr\_\To \tInl \tPair{x_1 + 1}{x_2 +y}})\end{array}\Bigg)}
{x}{\tPair{0}{0}},$$
where $c$ is a cut-off for floating-point underflow.

Next, recursive neural networks \cite{tai2015improved} are often mentioned as a use case of AD applied to recursive programs.
While basic Child-Sum Tree-LSTMs can also be implemented with
primitive recursion (a fold) over an inductively defined tree (which can be defined as a recursive type), there are other related models such
as Top-Down-Tree-LSTMs that require an iterative or general recursive approach \cite{zhang2016top}.
In fact, \cite{jeong2018improving} has shown that a recursive approach is preferable as it better
exposes the available parallelism in the model.
In the extended version of this paper \cite{LUCATELLI-VAKAR-2022-Extended}, we show some Haskell code for the recursive neural network of \cite{socher2011parsing}, to give an idea of how iteration and recursive types (in the form of inductive types of labelled trees) naturally arise in a functional implementation of such neural net architectures.
We imagine that many more applications of AD applied to recursive programs with naturally emerge as the technique made available to machine learning researchers and engineers.
Finally, we speculate that coinductive types like streams of real numbers, which can be encoded
using recursive types as $\mu \alpha.\Unit \To (\reals* \alpha)$, provide a useful API for on-line machine learning applications \cite{shalev2012online}, where data is processed in real time as it becomes~available.




\section{Categorical models for languages with recursive types}\label{sec:categorical-models}
We assume familiarity with basic category theory (see, for instance, \cite{MR0280560}).
We establish a class of categorical models for call-by-value (CBV) languages with tuple, variant, function, and recursive types, which we call \emph{$rCBV$ models}.

The first step is to establish the categorical model of computational $\lambda_C $-calculus (see \cite{moggi1988computational}). This means that the underlying structure is that of a \emph{Freyd-category}, see \cite{levy2003modelling}. 
Notwithstanding that, we do not need to consider this level~of generality. We call a pair  $\left( \catV , \monadT \right) $, 
where $\catV $ is bicartesian closed and $\monadT $ is a $\catV$-enriched monad,
a \emph{$CBV$ pair}. In this setting, we call $\catV $ the \emph{category of values}
and the corresponding Kleisli $\catV$-category $\catC $ the \emph{category of computations}. 

A \emph{$CBV$ pair morphism} between $CBV$ pairs $\left( \catV , \monadT \right) $ and $\left( \catV ', \monadT  \right) $
consists of a strictly bicartesian closed functor $H: \catV\to \catV '$ such that
$H \ee  = \ee ' _H $ and $ H \mm = \mm '_ H  $, where $\ee , \ee '$ and $\mm , \mm ' $ are the respective units and multiplications of $\monadT , \monadT '$.

\subsection{Parametric types and type recursion}
Let  $\left( \catV , \monadT \right) $ be a $CBV$ pair. For each $n\in\NN $, we can model $\left( n+1 \right) $-variable $\left( \catV , \monadT \right) $-parametric types as pairs $\left( \pE{E}{\catV}, \pE{E}{\catC} \right) $ of $\catV$-enriched functors such that \eqref{eq:parametric-types-in-terms-of-functors}  commutes, where $J $ is corresponding universal Kleisli $\catV$-functor from $\catV$ into the Kleisli $\catV$-category $\catC$.
The $0$-variable parametric types \eqref{eq:0-variable-parametric-types} are identified with objects of $\catV$. 
\\[-3pt]
\small \noindent\begin{minipage}{.5\linewidth}
	\begin{equation}\label{eq:parametric-types-in-terms-of-functors} 
		\diag{obvious-diagram-parametric-types-n}
	\end{equation}	
\end{minipage}%
\begin{minipage}{.5\linewidth}
	\begin{equation}\label{eq:resulting-recursive-parametric-type}
		\diag{obvious-diagram-parametric-types-(n-1)-recursive}
	\end{equation} 
\end{minipage}\normalsize\\[-3pt]

We model \textit{type recursion} in this setting. 
\textit{Let $\ParamAll{\catV}{\monadT}$ be the collection of all  $\left(\catV , \monadT \right) $-parametric types.}
A \textit{free type recursion} for $\left( \catV , \monadT \right) $ is a pair $\ffixpointRecT = \left( \fixpointRecT , \rolling \right)  $, where: \textbf{(1)} $\fixpointRecT$ is an operator that is the identity on $0$-variable parametric types and associates each $\left( n+1\right)$-variable parametric type  \eqref{eq:parametric-types-in-terms-of-functors} with an $n$-variable parametric type \eqref{eq:resulting-recursive-parametric-type}; \textbf{(2)} $\rolling$ is a collection \eqref{eq:rolling} of natural transformations such that  \eqref{eq:diag-roll}
is invertible: namely, $J\left(\rollReT ^{\pEE{E}} \right) $   is a natural isomorphism.
\\[-3pt]
\small \noindent\begin{minipage}{.4\linewidth}
	\begin{equation}\label{eq:0-variable-parametric-types}
		\left( \left( \catV ^\op \times \catV\right) ^0 \to \catV, \left( \catC ^\op \times \catC\right) ^0 \to \catC\right) 
	\end{equation}	
	\begin{equation}\label{eq:rolling}
		\rolling = \left( \rollReT ^{ \pEE{E} } \right) _ {\pEE{E}= \left( \pE{E}{\catV}, \pE{E}{\catC} \right)\in \ParamAll{\catV}{\monadT}  }  
	\end{equation} 	
\end{minipage}%
\noindent\begin{minipage}{.6\linewidth}
	\begin{equation}\label{eq:diag-roll}
		\diag{roll-basic-diagram}
	\end{equation} 
\end{minipage}\normalsize\\[-10pt]
A $CBV$ pair endowed with a free type recursion is our basic definition of model for our language with tuple, variant, function and recursive types.

\begin{definition}[$rCBV$ model morphism] 
	An \textit{$rCBV$ model} is a triple $\left(\catV , \monadT, \ffixpointRecT\right) $
	where $\left(\catV , \monadT\right) $ is a $CBV$ pair and $\ffixpointRecT$ is a free type recursion for $\left(\catV , \monadT\right) $. 
	
	An \textit{$rCBV$ model morphism} between $rCBV$ models $\left(\catV , \monadT, \ffixpointRecT \right) $ and $\left(\catV  ', \monadT ', \ffixpointRecT '\right) $ consists of a $CBV$ pair morphism $H: \catV\to \catV '$ such that, for each $n\in\NN$ and each pair	$\left( \pEE{E}, \pEE{E'}\right) $ of $\left( n+1\right) $-variable parametric types satisfying $\pE{ E'}{\catV '}\circ \left( H ^\op\times H \right) ^{n+1} = H\circ  \pE{ E}{\catV } $, we have that
	$\pE{\fixpointRecT E'}{\catV '}\circ \left( H ^\op\times H \right) ^{n} = H\circ  \pE{\fixpointRecT E}{\catV } $ and $H\left(\rollReT ^{\pEE{E}} \right) =\rollReT ^{\pEE{E}} _{\left( H^\op\times H \right) ^{n} } $.
	\textit{We denote by $\RCBVcat$ the category of $rCBV$ models and $rCBV$ model morphisms.} 
\end{definition}  


\subsection{Concrete models based on \wcpo{s}}
Herein, for simplicity's sake, we avoid the generality of \textit{bilimit compact expansions} by considering a subclass of concrete models, the  $rCBV$ $\wCpo$-pairs.\footnote{See \cite[4.2.2]{levy2012call} or \cite[Sect.~8]{vakar2020denotational} for the general setting of bilimit compact expansions.}

Let us write $\wCpo$ for the usual category of $\omega$-complete partial orders and monotone $\omega$-continuous functions. 
Recall that it is a complete, cocomplete cartesian closed category.
An $\wCpo$-category $\catV $ is \textit{$\wCpo$-cartesian closed} if $\catV $ has finite $\wCpo $-products and, moreover, for each object $B\in \catV$, the $\wCpo$-functor $\left(B\times -\right) $ has a right $\wCpo$-adjoint $\ihomV{B}{-}$.

A morphism \textit{$j$ in an $\wCpo$-category $\catB$ is full} if $\ehom{\catC}{B}{j}$ is a full morphism in $\wCpo$ for any $B\in\catC$. Moreover, an \textit{embedding-projection-pair (ep-pair)}  $u : A\epto B$
in an $\wCpo$-category $\catC$ 
is a pair $u = \eppair u$ consisting of a $\catC $-morphism
$\emb u:A\to B$, the \emph{embedding}, and a
$\catC$-morphism $\prj u:B\to A$, the \emph{projection}, such that
$\emb u \circ \prj u \leq \id$ and $\prj u\circ \emb u= \id$.
A zero object\footnote{Recall that a \emph{zero object} is an object that is both initial and
	terminal.} $\epzeroo$ in an $\wCpo$-category $\catC $ is an \emph{ep-zero object}  if, for any object $A$, the pair $\uniqMor _A = \left( \emb \uniqMor : \epzeroo\to A , \prj \uniqMor : A\to\epzeroo \right)  $ consisting of the unique morphisms is an ep-pair. 

\begin{definition}[$rCBV$ $\wCpo$-pair]\label{def:concrete-rCBV-models}
	An 	$rCBV$ $\wCpo$-pair is a $CBV$ pair $\left( \catV , \monadT \right) $ such that, denoting by $J : \catV\to\catC  $ the corresponding universal Kleisli $\catV$-functor,
	\begin{enumerate}[r1.]
		\item $\catV$ is a cocomplete $ \wCpo $-cartesian closed category;\footnote{$\catV$ is, hence, $\wCpo$-cocomplete as well.}\label{condition-for-rCBVwCpo-cocomplete}
		\item the unit of $\monadT$ is pointwise a full morphism (hence, $J$ is a locally full $\wCpo$-functor);\label{condition-for-rCBVwCpo-unit}
		\item  $\catC $ has an ep-zero object $\epzeroo = J\left( \initiall\right) $, where 
		$\initiall$ is initial in $\catV$;	\label{condition-for-rCBVwCpo-ep-zero}
		\item whenever $u : J(A)\epto J(B)$ is an ep-pair in $\catC $, there is one morphism 
		$\unlifteppair{u} : A\to B $ in $\catV$ such that $J\left( \unlifteppair{u}\right) =\emb u  $.\label{eq:pulling-embeddings-values}
	\end{enumerate}
	An  \textit{$rCBV$ $\wCpo$-pair morphism} from $\left(\catV , \monadT \right) $ into $\left(\catV  ', \monadT ' \right) $ is an $\wCpo$-functor 
	$H : \catV\to\catV ' $ that strictly preserves $\wCpo$-colimits, and whose underlying functor is a $CBV$ pair morphism.
	This defines a category of $rCBV$ $\wCpo$-pairs, denoted herein by $\wrCBVcat$.
\end{definition}
Every $rCBV$ $\wCpo$-pair $\left( \catV , \monadT \right) $ has an underlying 
$rCBV$ model. Namely, we have a canonical free type recursion $\wffixpointRecT = \left( \wfixpointRecT , \pwrollReT{E}_A \right) $
where $\wfixpointRecT$ is
constructed out of (bi)limits of chains of ep-pairs (see \cite[Section~8]{LUCATELLI-VAKAR-2022-Extended}, the extended version of the paper). 
This construction extends to a functor $\rCBVU : \wrCBVcat\to \RCBVcat $, which shows how
our concrete models are indeed $rCBV$ models.  

\subsection{Syntax as freely generated $rCBV$ models}\label{subsect:awesome-subsection-about-freely-fenerated-categ-structures}
In Section \ref{sec:AD-macro}, we will consider the syntax of an ML-family programming language with recursive types.
It is generated from certain \emph{primitive types} (like a type $\reals$ for real numbers)
and certain \emph{primitive operations} $\op$ (e.g. mathematical operations like $\sin,\cos, \exp,(+),(*)$, etc.).
We can consider the syntax of our languages as $rCBV$ models by taking the types $\tau$ of the language as objects and equational equivalence classes  of programs $x:\tau\vdash t:\sigma$ as morphisms $\tau\to \sigma$.
This is a freely generated $rCBV$ model in the sense that we get a unique $rCBV$ model morphism 
to any $rCBV$ model once we fix the image of all primitive types and operations in a consistent way. We call these \textit{freely generated $rCBV$ models  on a language} \textit{syntactic $rCBV$ models}.

\section{Subscone}\label{sec:LR}
We establish, here, the basic categorical framework underlying the logical relations \textit{(LR)} argument. 
Our general view is that the categorical approach to semantic logical relations relies on studying principled ways to construct concrete categorical semantics out of elementary ones. This construction should be informed of the desired properties to be proved: so that the resulting semantics assures us of the property we want to establish in each setting.

The first step is to choose a basic concrete categorical semantics for our
language. For $CBV$ languages with recursive types like ours,  a  \textit{basic} concrete model usually consists of an $rCBV$ $\wCpo$-pair. We are particularly interested in the elementary $rCBV$ $\wCpo$-pair  $\left( \wCpo ^n , \monadwP{-} \right) = \left( \wCpo  , \monadwP{-} \right) ^n  $ for some $n\in \NN$, where $\monadwP{-}$ is the usual partiality $\wCpo$-monad that freely adds a least element $\leastelement $ to each \wcpo{}. 

It is tempting to add or consider more structured semantics and appeal to more general theories. However, we believe that adding structure beforehand is an \textit{ad hoc} anticipation of constructing our semantic logical relations proof, which is avoidable if we have powerful enough principled techniques.

Given a chosen basic $rCBV$ $\wCpo$-pair $\left(\catV , \monadT\right) $, with a well defined semantics $\sem{-} $ for our language, 
the second step is to establish the base LR $\wCpo$-functors. These are suitable $\wCpo$-functors 
$G: \catV \to \wCpo $ such that we can express the property we want to prove starting from predicates over $G\left(Z\right)\in \wCpo $. 

In this direction, in the context of semantic open logical relations (and in our $AD$ setting), we often want to consider the family of $\wCpo$-functors given by $\left( \catV\left( \sem{\ty{1}} , -\right) : \catV\to \wCpo  \right)_{\ty{1}\in \PTYPE} $ indexed by a subset of types $\ty{1} $ of the language.  

Given such a family of $\wCpo$-functors $\left( G _{\ty{1}}:\catV\to\wCpo\right) _{\ty{1}\in \PTYPE}$, we consider the $\wCpo$-scone along $G _{\ty{1}}$ (for each $\ty{1}$): these are the comma $\wCpo$-categories $\wCpo\downarrow G _{\ty{1}}$ whose definition we recall below.
\begin{itemize} 
	\renewcommand\labelitemi{--}	
	\item  The objects of $\wCpo\downarrow G _{\ty{1}}$ are triples $(D\in\wCpo  , C\in\catV  , j:D\to G(C) )$ in which  $j$ is a morphism of $\wCpo$; 
	\item  a morphism $(D,C, j)\to (D', C', h)$ between objects of $\wCpo\downarrow G _{\ty{1}}$ is a pair \eqref{eq:pair-alpha-scone} making \eqref{eq:comma-morphism-scone} commutative in $\catD $;
	\item if \eqref{eq:pair-alpha-scone} and \eqref{eq:pair-beta-scone}
	are two morphisms in $\wCpo\downarrow G _{\ty{1}}\left( \left( D, C , j\right), \left( D', C', h\right) \right) $, 
	we have that $\alpha\leq \beta $ if $\alpha _0\leq  \beta _0 $ in $\wCpo $ and $\alpha _1\leq \beta _1$ in $\catV $.
\end{itemize} 
\begin{minipage}{.5\linewidth}
	\small	\begin{equation}\label{eq:pair-alpha-scone}
		\alpha = \left( \alpha_0 : D\to D'  , \alpha _1 : C\to C'\right)
	\end{equation} 		
	\begin{equation}\label{eq:pair-beta-scone}
		\beta = \left( \beta _0 : D\to D'  , \beta _1 : C\to C'\right) 
	\end{equation}
\end{minipage}%
\begin{minipage}{.5\linewidth}
	\normalsize	\begin{equation}\label{eq:comma-morphism-scone}
		\diag{morphism-comma-category}
	\end{equation}   
\end{minipage} \\\normalsize
Provided that $G_{\ty{1}} $ is a right $\wCpo$-adjoint, the forgetful $\wCpo$-functor $ \forgetfulS _{\ty{1}} : \wCpo \downarrow G_{\ty{1}}\to \wCpo\times\catV $ is $\wCpo$-comonadic and $\wCpo$-monadic. We can conclude, then, that it creates (and strictly preserves) $\wCpo$-colimits and limits. Moreover, $\wCpo \downarrow G _{\ty{1}}$ is $\wCpo$-bicartesian closed.
This is the $\wCpo$-enriched version of the results presented in \cite[Section~9]{VAKAR-LUCATELLI2021}
(see \cite[Appx.~C]{LUCATELLI-VAKAR-2022-Extended}).

In order to proceed with a proof-irrelevant approach, we consider the subscone: namely,  
$\SUBscone{\catD}{G_{\ty{1}}}$ is, herein, by definition the full $\wCpo$-subcategory of 
$\wCpo \downarrow G _{\ty{1}}$ whose objects are triples $\left( D\in\catV , C\in\wCpo, j\right)$ where $j$ is a full morphism. $\SUBscone{\catD}{G_{\ty{1}}}$  is, then, a full reflective and replete
$\wCpo$-subcategory of $\wCpo\downarrow G _{\ty{1}}$. Moreover, we have that:
\begin{theorem}\label{theo:main-subscone-theorem-total}
	$\SUBscone{\wCpo}{G _{\ty{1}}}$ is cocomplete and $\wCpo$-cartesian closed. Moreover, the forgetful $\wCpo $-functor $\forgetfulSub _{\ty{1}} : \SUBscone{\wCpo}{G_{\ty{1}}}\to \catV $ is strictly
	$\wCpo$-colimit preserving, cartesian closed and locally full (hence, faithful). 
\end{theorem} 	
The reader interested in further considerations on the result above may take a look at \cite[Section~6]{LUCATELLI-VAKAR-2022-Extended}.
Theorem \ref{theo:main-subscone-theorem-total} shows how $\SUBscone{\wCpo}{G _{\ty{1}}}$ already yields a model for the category of values of our language. The remaining step consists in giving the conditions under which  a  $\SUBscone{\wCpo}{G _{\ty{1}}}$-monad $\LRMONAD{\ty{1}}$ yields an $rCBV$ $\wCpo$-pair and
an $rCBV$ $\wCpo$-pair morphism from $\left( \SUBscone{\wCpo}{G _{\ty{1}}}, \LRMONAD{\ty{1}} \right) $ into  $\left( \catV , \monadT\right) $.

\subsection{Recursive types for lifted monads}\label{subsect:LR-as-rCBV-model}
In the context of Theorem \ref{theo:main-subscone-theorem-total}, there might be canonical/universal liftings of the monad $\monadT$ along $\forgetfulSub _{\ty{1}} $. However, these are not necessarily the monads we want -- since we are assuming that we started out from a very basic semantics, defined in an $rCBV$ $\wCpo$-pair $\left( \catV , \monadT\right)  $. In this scenario, our lifting should be informed with the logical relations we want for the computations of primitive types. 

Let $\LRMONAD{\ty{1}}$ be a strong $\SUBscone{\wCpo}{G _{\ty{1}}}$-monad that is a lifting of $\monadT $ along $\forgetfulSub _{\ty{1}} $. This means that $\forgetfulSub _{\ty{1}} $ yields a $CBV$ pair morphism \eqref{eq:rCBV-model-morphism-subsconing}. Assume that  $\LRMONAD{\ty{1}}$ satisfies the following two properties: \textbf{(A)} for each object $\left( D, C, j \right) $ of $\SUBscone{\wCpo}{G _{\ty{1}}}$, the square 
induced by each component of the unit of $\LRMONAD{\ty{1}}$ in $\wCpo$ is a pullback; and \textbf{(B)} $\LRMONAD{\ty{1}}\left( \initiall , \initiall, j\right) $ is the terminal object in $\SUBscone{\wCpo}{G _{\ty{1}}}$. In this setting, we get that Theorem \ref{theo:subsconing-generic} holds (by the main result established in \cite[8.8]{LUCATELLI-VAKAR-2022-Extended}).

\begin{theorem}\label{theo:subsconing-generic}
	$\left( \SUBscone{\wCpo}{G _{\ty{1}}}, \LRMONAD{\ty{1}}\right)$ is an $rCBV$ $\wCpo $-pair. Moreover,
	\eqref{eq:rCBV-model-morphism-subsconing} 
	yields an $rCBV $ $\wCpo $-pair morphism. 
	\begin{equation}\label{eq:rCBV-model-morphism-subsconing}
		\forgetfulSub _{\ty{1}}: \left( \SUBscone{\wCpo}{G _{\ty{1}}}, \LRMONAD{\ty{1}}\right)\to \left( \catV , \monadT\right) 
	\end{equation}
\end{theorem} 
In our use case of Theorem~\ref{theo:subsconing-generic}, 
the remaining step to define the logical relations is to define a morphism  $\SEMLRR{\op }{\ty{1} }\in\SUBscone{\wCpo}{G _{\ty{1}}} $
for each primitive operation $\op $ in a compatible way. This yields an $rCBV$ model morphism $\SEMLRR{ - }{\ty{1} } $
from the free $rCBV $ model on the syntax of our language to  $\rCBVU\left( \SUBscone{\wCpo}{G _{\ty{1}}}, \LRMONAD{\ty{1}}\right)$.

\section{Syntax and AD macro for an ML-like language}\label{sec:AD-macro}
We establish basic example of languages with type recursion where we can do automatic differentiation. Since it relates better to the efficient implementations we have in mind~\cite{smeding2022}, we see \textit{automatic differentiation} (AD) as a program transformation between two languages. The target language is a simple extension of the source language: we add an extra type $\tangentreals $ of (co)tangent vectors.

\subsection{Source language}
For the source language, we consider a standard (coarse-grain) call-by-value language with ML-style polymorphism and type recursion in the sense of FPC \cite{fiore1994axiomatisation}. The language is constructed over a ground type $\reals$, certain real constants $\cnst{c}\in\Op_0$,
certain primitive operations $\op\in\Op_n$ for each nonzero natural number $n\in\NN ^\ast $, and $\tSign$.
We denote $\displaystyle\Op := \bigcup _{n\in\NN } \Op _ n $.

$\reals $ intends to implement floating point or exact real numbers. Moreover,
for each $n\in\NN$, the operations in $\Op _n $ intend to implement partially defined functions
$\RR ^n \rightharpoonup \RR $. Finally, $\tSign $ intends to implement the partially defined function $\signR :\RR\rightharpoonup \RR$ defined in $\RR ^- \cup \RR ^+ $
which takes $\RR ^-$ to $-1$ and $\RR ^+ $ to $1$.


We will later interpret these operations as morphisms $\RR^n\to \monadwP{\RR}$ in $\wCpo$ with domains
of definition that are (topologically) open in $\RR^n$, on which they are differentiable functions.
These operations include, for example, unary operations on reals like $\exp,\log,\sigmoid\in \Op_1$ (where we mean the mathematical sigmoid function
$\sigmoid(x)\defeq\frac 1 {1+e^{-x}}$),
binary operations on reals like $(+),(-),(*),(/)\in\Op_2$, or any other desired partially defined differentiable function $\RR^n\to \monadwP{\RR}$ (see Sect. \ref{sec:semantics-of-differentiation}).

We treat these operations in a schematic way as this reflects the reality of 
practical AD libraries, which are constantly being expanded with new primitive operations.
The types $\ty{1},\ty{2},\ty{3}$, values $\val{1},\val{2},\val{3}$, and computations $\trm{1},\trm{2},\trm{3}$ of our language are as follows - they are standard:\footnotesize\\
\footnotesize
\begin{syntax}
    \ty{1}, \ty{2}, \ty{3} & \gdefinedby\!\!\! & & \syncat{types}                          \\
    &\gor& \reals                      & \synname{numbers}\\
    &\gor & \Init  \gor \ty{1} \t+ \ty{2}  & \synname{sums}\\
    &\gor\; & \Unit  \gor  \ty{1}_1 \t* \ty{1}_2 & \synname{products}  
  \end{syntax}
 \begin{syntax}
  &\gor& \ty{1} \To \ty{2}              & \synname{function}      \\
    &\gor & \tvar{1},\tvar{2},\tvar{3}   & \synname{type variables}\\
  &\gor\; & \trec{\tvar{1}}\ty{1} & \synname{recursive type}
\end{syntax}\\
\footnotesize
\begin{syntax}
	\val{1}, \val{2}, \val{3} & \gdefinedby\!\!\! & & \syncat{values}                          \\
	&\gor& \var{1},\var{2},\var{3}                      & \synname{variables}\\
	&\gor& \cnst{c}                   & \synname{constants}\\
	&\gor& \tInl{\val{1}} \gor   \tInr{\val{1}} & \synname{inclusions}\\
	&&&\\
	\trm{1}, \trm{2}, \trm{3} & \gdefinedby\!\!\! & & \syncat{computations}                          \\
	&\gor& \var{1},\var{2},\var{3}                      & \synname{variables}\\
	&\gor & \letin{\trm{1}}{\var{1}}{\trm{2}} & \synname{sequencing}\\
	&\gor& \cnst{c}                   & \synname{constant}\\
	&\gor & \op(\trm{1}_1,\ldots,\trm{1}_n) & \synname{operation}\\
	&\gor& \tInl{\trm{1}} \gor   \tInr{\trm{1}} & \synname{inclusions}\\
	&\gor & \vMatch{\trm{3}}{\begin{array}{l}\;\;\tInl\var{1}\To\trm{1}\\
			\gor \tInr\var{2}\To \trm{2}\end{array}}\hspace{-17pt}\; & \synname{sum match}\\
\end{syntax}
\begin{syntax}
	&\gor\; & \tUnit \ \gor  \tPair{\val{1}}{\val{2}} & \synname{tuples}\\
	&\gor& \fun{\var{1}}{\trm{1}}             & \synname{abstractions}      \\
	&\gor& \tRoll\val{1}                      & \synname{recursive intro}\\
	&&&\\\\
	&\gor & \nvMatch{\trm{1}} & \synname{sum match}\\
	&\gor\; & \tUnit \ \gor  \tPair{\trm{1}}{\trm{2}} & \synname{tuples}\\
	&\gor\; & \pMatch{\trm{2}}{\var{1}}{\var{2}}{\trm{1}}\hspace{-10pt}\; & \synname{product match}\\
	&\gor& \fun{\var{1}}{\trm{1}}             & \synname{abstractions}      \\
	&\gor& \trm{1}\ \trm{2}             & \synname{function app.}      \\
	&\gor&\tSign\trm{1} & \synname{sign function}\\
	&\gor& \tRoll\trm{1}                      & \synname{recursive intro}\\
\end{syntax}
\normalsize\\
\normalsize We use the sugar  
$\ifelse{\trm{3}}{\trm{1}}{\trm{2}}\defeq 
\vMatch{\tSign\trm{3}}{{
		\_\To{\trm{1}}
		\vor \_\To{\trm{3}}
}}$ for lazy real conditionals.

The typing rules for computations are standard, with the exception of the typing of constants and primitive operations which are listed in Fig.~\ref{fig:type-system-source-short}.
\begin{figure}[!t]
	\fbox{
		\parbox{0.98\linewidth}{
			\input{type-system-source-short}}}
	\caption{\label{fig:type-system-source-short} The type assignment rules for the primitive type of $\reals$.}
\end{figure}
(We list the full rules in Fig.~\ref{fig:typestermrecursion-iteration} in App.~\ref{app:Typing-Rules}.) We consider the standard CBV $\beta\eta$-equational theory (see \cite{moggi1988computational}) for our language (which we list in Fig. \ref{fig:beta-eta}).

\subsection{Target language} 
As mentioned above, our target language is a simple extension of our source language, introducing a new type $\tangentreals$ that plays the role of (co)tangents.  
We do so by adding the following syntax, with the typing rules of
Fig.~\ref{fig:type-system-target}
(for the full typing rules, see Fig.~\ref{fig:types-target-cotangent} in App.~\ref{app:Target-Typing-Rules}). 
\\
\begin{syntax}
    \ty{1}, \ty{2}, \ty{3} & \gdefinedby & & \syncat{types}                          \\
    &\gor& \ldots                      & \synname{as before}\\
    \val{1},\val{2},\val{3} & \gdefinedby && \syncat{values}\\
&\gor&\ldots&\synname{as before}\\
   &\gor &\cncanoni{i}&\synname{$i$-th canonical element} \\
   &\gor  & \cnzero & \synname{zero}\\
  \end{syntax}%
~
 \begin{syntax}
  &\gor\quad\, & \tangentreals  & \synname{(co)tangent}\\
\\
&\gor & \trm{1} + \trm{2}&\synname{addition of vectors}\\
&\gor & \trm{1}\ast\trm{2} & \synname{scalar multiplication} \\ 
  & \gor & \tangentprojection{i}{\trm{1}} & \synname{proj. handler}
\end{syntax}\\
\begin{syntax}
  \trm{1},\trm{2},\trm{3} & \gdefinedby && \syncat{computations}\\
  &\gor&\ldots&\synname{as before}\\
&\gor&\cncanoni{i}&\synname{canonical element}\qquad\;\;\\
&\gor & \cnzero & \synname{zero }\\\end{syntax}
~\begin{syntax}
  &\gor\quad & \trm{1} + \trm{2}&\synname{addition of vectors}\\
  &\gor & \trm{1}\ast\trm{2} & \synname{scalar multiplication} \\ 
    & \gor & \tangentprojection{i}{\trm{1}} & \synname{proj. handler}
\end{syntax}\\[-12pt] 
\begin{figure}[!h]\fbox{\parbox{0.98\linewidth}{
			\input{type-system-target-no-kinds}}}
	\caption{\label{fig:type-system-target} The type assignment rules for the $\tangentreals$ type of cotangents.}
\end{figure}

We want $\left( \tangentreals , + , \ast , \cnzero\right) $  to implement the vector space $\left( \RR ^k, + , \ast , 0\right) $, for some $k\in\NN\cup\left\{ \infty \right\}$\footnote{$\RR ^ \infty $ is the vector space freely generated by the infinite set $\left\{ \canonicalbasise{}{i}: i\in\NN  ^\ast \right\} $. In other words, it is the infinity coproduct of $\RR ^i $. In order to implement it, one can use lists/dynamically sized arrays and pattern matching for the vector addition.}.  In this case:  \textbf{(A)} $\cncanoni{i}$ should implement the $i$-th element  $\canonicalbasise{k}{i}\in\RR ^k $  of the canonical basis if $k=\infty $ or if $i\leq k $, and $0\in\RR^k$ otherwise; \textbf{(B)} $\tangentprojection{i}{\trm{1}}$ should
implement $\semanticshandler{k}{i}: \RR ^k \to \RR ^i$ 
which denotes the canonical projection if $i\leq k $ and the coprojection otherwise.

For short, we say that \textit{$\tangentreals $ implements the vector space $\RR ^k $} to refer to the case above.
For efficient implementations~\cite{smeding2022}, the operational semantics should make clever use of the usual distributive law of the scalar multiplication over the vector addition (aka linear factoring \cite{brunel2019backpropagation}). Although this is negligible from our semantical viewpoint, we observe that this optimisation is semantically correct.


\subsection{Languages as rCBV models}\label{sect:languages-as-rCBV-models}
As stressed in \ref{subsect:awesome-subsection-about-freely-fenerated-categ-structures}, we
have the \textit{syntactic $rCBV$ models} given by the freely generated  $rCBV$ models on our source and target languages, respectively denoted herein by
$rCBV$ models $\left( \SynVr, \SynTr, \SynffixpointRecT   \right)$ and $	\left( \SynVrt , \SynTrt ,\SyntffixpointRecT  \right)$.

\begin{theorem}[Universal property of syntax]\label{theo:section-universal-property-syntax}
	Let $ \left(\catV , \monadT, \ffixpointRecT \right) $   be an $rCBV$ model. Assume that \eqref{fig:assignment-functor-universalproperty-syntax} and \eqref{fig:assignment-functor-universalproperty-syntax-target} are given consistent assignments for each $\op\in\Op$, $\cnst c $ and $i\in\NN ^\ast$. There are unique $rCBV$ model morphisms\\ $H:\left( \SynVr, \SynTr, \SynffixpointRecT   \right)\to  \left(\catV , \monadT, \ffixpointRecT \right) $ respecting \eqref{fig:assignment-functor-universalproperty-syntax} and \\	 	
	$\extendedH{H}: \left( \SynVrt , \SynTrt ,\SyntffixpointRecT  \right)\to \left(\catV , \monadT, \ffixpointRecT \right)  $, where $\extendedH{H}$ extends $H$ and respects \eqref{fig:assignment-functor-universalproperty-syntax-target}.
	\begin{eqnarray}
		\small	\left( H(\reals)\in	
		\obb{\catV}, H (\tSign) , 
		H (\cnst c) , 
		H (\op)\right) \label{fig:assignment-functor-universalproperty-syntax} \\
		\left( \extendedH{H}( \tangentreals ) \in   \obb{\catV}, \extendedH{H} (\cnzero ), \extendedH{H}\left( \tangentprojection{i}{}\right), \extendedH{H} (+) , \extendedH{H} (\ast )\right)  \label{fig:assignment-functor-universalproperty-syntax-target}
	\end{eqnarray}\normalsize
\end{theorem}
\subsection{Dual numbers AD code transformation}
We define our \textit{automatic differentiation macro} by making use of the universal property of the source language. 
Let us fix, for all $n\in\NN$,  $\op\in\Op_n$, and  $1\leq i \leq n$, 
computations $\var{1}_1:\reals,\ldots,\var{1}_n:\reals\vdash \partial_i\op(\var{1}_1,\ldots,\var{1}_n):\reals$,
which represent the partial derivatives of $\op$.

The assignment defined in Fig.~\ref{fig:assignment-AD-functor} induces a unique $rCBV$ model morphism \eqref{eq:macro-as-a-functor-recursive-types}, 
which corresponds to the structure preserving macro $\Dsynsymbol$ on the types and computations of our language for performing AD defined in Fig.~\ref{fig:MACROAD}. 
\begin{equation}\label{eq:macro-as-a-functor-recursive-types}
	\DSynrec : \left( \SynVr, \SynTr, \SynffixpointRecT   \right) \to \left(\SynVrt{,}\SynTrt{,}\SyntffixpointRecT\right).\vspace{-12pt}
\end{equation} 

\begin{figure}[!t]
	\fbox{\parbox{0.98\linewidth}{\begin{minipage}{\linewidth}\noindent\input{SHORTAssignment-AD-transformation}\end{minipage}}}
	\caption{AD assignment, for each primitive operation  $\op\in\Op_n$ ($n\in\NN$) and each constant  $c\in \RRsyntax$.
		\label{fig:assignment-AD-functor}}
\end{figure} 	
\begin{figure}[!ht]
	\fbox{\parbox{0.98\linewidth}{\begin{minipage}{\linewidth}\noindent
				\input{d-types1SHORT}
				\hrulefill
				\input{d-terms1SHORT}
	\end{minipage}}}
	\caption{AD macro  $\Dsyn{-}$ defined on types and computations.
		All newly introduced variables are chosen to be fresh.}\label{fig:MACROAD}
\end{figure}

\section{Semantics of dual numbers AD}\label{sec:semantics-of-differentiation} 
Dual numbers AD relies on the interleaving of the function with its derivative. Semantically, we do so by considering the interleaving morphism between
discrete \wcpo{s}. We briefly establish the formal definitions in this section.
\textit{Henceforth, unless stated otherwise, the cartesian spaces $\RR ^n$ and its subspaces are endowed with the discrete $\wCpo $-structures, i.e. $x\leq y$ iff $x=y$, for all $x,y\in \RR ^n$.} 

\begin{definition}[Interleaving function]
	For each $(n,k)\in\NN \times \left( \NN\cup\left\{ \infty \right\}\right)  $, denoting by $\NNN{n}$ the set $\left\{ 1,\ldots , n\right\}$, we define the isomorphism (in $\wCpo$)
	\begin{eqnarray}
		\intle{n, k} : & \RR ^n\times \left( \RR ^k\right) ^n & \to \left( \RR \times \RR ^k \right) ^{n}\\
		&\left( (x_j)_{j\in\NNN{n}}, (y_j)_{j\in\NNN{n}} \right)  & \mapsto  \left( x_j, y_j \right) _{j\in\NNN{n}}.\nonumber 
	\end{eqnarray}
	For each open subset $U\subset \RR^n$, we denote by $\intle{n, k}^U : U\times \left( \RR ^k\right) ^n \to  \intle{n,k}\left( U\times\left( \RR ^k\right) ^n \right) $
	the isomorphism obtained from restricting $\intle{n,k} $. 
\end{definition} 

Before we proceed to define the interleaved derivative of a total defined function, we have two remaining 
relevant remarks. Firstly, as opposed to the usual,  we also care about the  derivative of functions between (possibly infinite) coproducts of open subspaces 
of various dimensions. We extend the usual  definition for this case in the obvious manner: it is just the corresponding definition for the free cocompletion of the category of 
connected spaces under coproducts.
Put differently, we treat derivatives a local operations on functions -- they can be computed restricted to each connected component of the input and glued together. The second remark is that we are particularly interested in a certain presentation of the derivative, via the transpose of the derivative (the natural semantical counterpart of reverse-mode AD). We present the precise definition.

\begin{definition}[Vectorised derivative]\label{def:total-derivative}
	Let $\displaystyle g : U\to \coprod _{j\in L} V_j  $ be a map where $U$ is an open subset of $\RR ^n $, and, for each $i\in L$, $V_i $ is an open subset of $\RR ^{m_i}$. 
	
	The map $g$ is \textit{differentiable} if, for any $i\in L$, $g^{-1}\left( V_i \right) =W _i $ is open in $\RR ^n$ and the restriction $g|_{W_i} : W_ i \to V_i $ is differentiable w.r.t the submanifold  structures $W_i\subset \RR ^{n}$ and $ V_i \subset \RR ^{m_i} $.		 
	In this case, for each $k\in  \left( \NN\cup\left\{ \infty \right\}\right) $, we define the function $\dDSemtotaltra{k}{g}$, which we think of as the $k$-dimensional vectorised derivative: 
	\begin{eqnarray}
		\dDSemtotaltra{k}{g} : & \intle{n, k}\left(  U\times\left( \RR ^k\right)  ^n\right)  &\to \coprod _{j\in L}\left( \intle{{m_j}, k}\left( V_i\times\left( \RR ^k\right)  ^{m_i}\right)\right)\label{eq:definition-of-total-derivative-k-interleaved} \\
		& z &\mapsto \coproje{m_j}\circ\intle{{m_j},k}^{V_j}\left( g(x), \vectoraslineartransformation{w}\cdot g'(x)^t\right) ,\nonumber\\
		&&  \text{whenever } \intle{n, k} ^{-1} \left( z \right) =\left( x, w \right)\in W_i\times \left(\RR ^k\right) ^n 		\nonumber
	\end{eqnarray}
	in which $\vectoraslineartransformation{w}$ is the linear transformation $ \RR ^n\to \RR ^k $ corresponding to the vector $w$, $\cdot $ is the composition of linear transformations, $\coproje{m_i}$ is the obvious $ith$-coprojection of the coproduct (in the category $\wCpo$),    and $g'(x)^t $ is the transpose of the derivative $g'(x) :\RR ^n\to\RR ^{m_i} $ of $g|_{W_i} : W_ i \to V_i $ at $x\in U$.
\end{definition}

We extend the definition above to the case of partially defined functions. More precisely,
a partially defined function given by  $ h_i  : \RR ^{n_i }\to \monadwP{ \coprod\limits_{j\in L } \RR ^{m_j}   }  $
\textit{is differentiable} if its domain of definition $U_i$  is open in $\RR ^{n_i }$, and the corresponding total function defined in $U_i$ is differentiable in the sense of Def.~\ref{def:total-derivative}. 
Moreover, in this case, we define the \textit{derivative $\dDSemtra{k}{h_i} $} by \eqref{eq:derivative-parially-defined-functions}. Finally, a partially defined function $  h : \coprod\limits_{r\in K } \RR ^{n_r} \to\monadwP{ \coprod\limits_{j\in L } \RR ^{m_j}   }  $ if  every component $\displaystyle h_i := h\circ \coproje{i}$ is differentiable. In this case, the derivative $\displaystyle\dDSemtra{k}{h} $ is defined by $ \cpairL \dDSemtra{k}{h_r } \cpairR _{r\in K}$. 
\begin{eqnarray}
	\dDSemtra{k}{h_i} : & \left( \RR \times \RR  ^k \right) ^{n_i} &\to\monadwP{\coprod_{j\in L } \left( \RR \times  \RR ^k  \right) ^{m_j} } \label{eq:derivative-parially-defined-functions}\\
	&z &\mapsto 
	\begin{cases}
		\dDSemtotaltra{k}{h_i}\left( z\right) , & \text{if } z\in \intle{{n_i}, k}\left(  U_i\times\left( \RR ^k\right)  ^{n_i}\right)\subset \left( \RR \times \RR  ^k \right) ^{n_i};
		\\
		\leastelement  , & \text{otherwise.} 
	\end{cases}\nonumber
\end{eqnarray}

\subsection{Basic semantics for the languages}\label{subsect:semantics-source-language}
We give a concrete semantics for our language, interpreting it in the $rCBV$ $\wCpo$-pair $\left( \wCpo , \monadwP{-}\right) $: namely, by Theorem \ref{theo:section-universal-property-syntax},
we get a unique $rCBV$ model morphism \eqref{eq:functor-semantics}  defined by the assignment of  $\reals $  and each primitive operation $\tSign $, $\cnst c$, $\op$ with  the corresponding intended semantics $\RR $ (with the discrete $\wCpo$-structure),  
$\sem{\tSign }\coloneq \signR : \RR \to \monadwP{\terminall \sqcup \terminall }$, $\sem{\cnst c}\coloneq \semanc{c}: \RR \to \monadwP{\terminall\sqcup \terminall }  $,  $\sem{\op}\coloneq\seman{f}: \RR ^n \to\monadwP{\RR  }  $.
Moreover, for each $k\in\NN \cup\left\{ \infty \right\} $, we can extend $\sem{-}$ into an $rCBV$ model morphism $\semt{k}{-}$ by interpreting $\left( \tangentreals ,  + , \ast, \cnzero \right)$ as the vector space $\left( \RR ^k,  + ,  \ast , 0 \right) $ it intends to implement, and the handlers as the appropriate (co)projections. The $rCBV$ model morphism $\semt{k}{-}$ defines the \textit{$k$-semantics for the target language}. We can, then, consider the $k$-combined semantics of the product of our languages:
\small \begin{equation} \label{eq:functor-semantics} 
	\sem{-}\times  \semt{k}{-} :\left( \SynV , \SynT ,\SynffixpointRecT  \right)\times \left( \SynVrt , \SynTrt ,\SyntffixpointRecT  \right) \to \CBVU\left( \wCpo ^2, \monadwP{-}\right) 
\end{equation} 	
\normalsize

We want to prove the semantical correctness of our macro $\Dsynsymbol $, defined by $\DSyn $: namely, 
denoting by $\dDSemj{f}{j}$ the usual $j$-th partial derivative of $f$, 
assuming that \eqref{eq:BASIC-ASSUMPTION}  holds for any primitive $\op\in\Op$, we want to get \eqref{eq:aim-semantical-result}  for any $t$ between objects
corresponding to data types in our language. It should be noted that, from \eqref{eq:BASIC-ASSUMPTION}, we already get \eqref{eq:trivial-consequence-of-soundness}.\footnotesize \\
\noindent\begin{minipage}{.40\linewidth}	
	\begin{equation}\label{eq:BASIC-ASSUMPTION} 
		\sem{\partial_j\op(\var{2}_1,\ldots,\var{2}_n)   } 
		= \dDSemj{\sem{\op } }{j}
	\end{equation} 	
\end{minipage}%
\noindent\begin{minipage}{.26\linewidth}
	\begin{equation}\label{eq:aim-semantical-result}
		\semt{k}{\Dsyn{ t  }} = \dDSemtra{k}{\sem{t}}
	\end{equation}
\end{minipage} 
\begin{minipage}{.26\linewidth}
	\begin{equation}\label{eq:trivial-consequence-of-soundness}
		\semt{k}{\DSyn{\left( \op\right) }} = \dDSemtra{k}{\sem{\op}}
	\end{equation}
\end{minipage}  \\
\normalsize

\section{Subscone for partially defined differentiable functions}\label{sec:logical-relations}
Henceforth, we assume that $\tangentreals$ implements the vector space $\RR ^k $, and that \eqref{eq:trivial-consequence-of-soundness} holds.
We establish the categorical framework of our correctness proof. Since we already established the basic semantics \eqref{eq:functor-semantics}, 
we start by establishing our subscone: more precisely, the family of base LR $\wCpo$-functors  satisfying the setting of  Section~\ref{sec:LR}.

The differentiability and the derivative of a total function $g: \RR ^m \to C $ (where $C$ is some manifold)
are fully characterized by the differentiability and the derivatives of  $g\circ \alpha $ for differentiable maps $\RR ^n \to \RR ^m $ ($n\in\NN $). More precisely, 
\textit{a total function $g: \RR ^m \to C $ is differentiable 
	and $\dot{g} = \dDSemtotaltra{k}{g} $ if and only if  $g\circ \alpha $ is differentiable and  
	$\dot{g}\circ \dDSemtotaltra{k}{\alpha}  = \dDSemtotaltra{k}{\left( g\circ\alpha \right)  } $
	for any differentiable map $\alpha: \RR ^n \to \RR ^m $ (and any natural $n$).} One direction of this observation follows from the \textit{chain-rule for derivatives}, while the other is trivial since we can take $\alpha = \ID $. 


This leads us to consider the family defined by \eqref{eq:family-of-base-LR}, which clearly satisfies the setting of Section~\ref{sec:LR}: hence, we have that, for each $n$, $ \SUBscone{\wCpo }{\sconeFUNCTOR{n} }$ is an appropriate model for the values of our languages by Theorem \ref{theo:main-subscone-theorem-total}.
\footnotesize \begin{equation} \label{eq:family-of-base-LR} 
	\left( \sconeFUNCTOR{n} \coloneq \ehom{\wCpo\times\wCpo}{\left( \RR^n , \left( \RR\times \RR^k\right) ^n  \right)}{  \left( - , - \right)  } : \wCpo\times \wCpo \to \wCpo\right) _{n\in\NN} 
\end{equation}  
\normalsize
We proceed to define a monad lifting satisfying the setting of \ref{subsect:LR-as-rCBV-model} in order to end up with an $rCBV$ $\wCpo$-pair and, hence, an $rCBV$ model. We do that informed of the property that we want to 
prove. 
We start by extending the observation about the characterization of differentiability to 
the case of partially defined functions: namely,   $  h :  \RR ^{n_i} \to\monadwP{ \coprod\limits_{j\in L } \RR ^{m_j}   }  $ is differentiable and $\dot{h} = \dDSemtra{k}{h_i} $ if and only if: \textbf{(A)} the domain of definition of $h$ is an open set $U_i\subset  \RR ^{n_i} $;
\textbf{(B)} for any differentiable map $\alpha: \RR ^n \to U_i $ and any $n\in\NN $, $h_{U_i}\circ \alpha $ is differentiable and, 
denoting by $\unlift{\dot{h}} $ the total function corresponding to $\dot{h}$, 
$\unlift{\dot{h}}\circ \dDSemtotaltra{k}{\alpha}$ is well defined and equal to $\dDSemtotaltra{k}{\left( h_{U_i}\circ\alpha \right)  } $.

Informed of this observation and in order to establish the underlying predicate above, we define, for each 
$U\in\topologyO{n}$, where $\topologyO{n}$ is the set of proper open non-empty subsets of the cartesian space $\RR^n $, the object $\openLift{U, n}$: 
\footnotesize
\begin{eqnarray*} 
	\openLift{U, n} &\!\defeqq \!&\left( \left\{ \left( g: \RR ^n\to U, \dDSemtotaltra{k}{g} \right) : g \mbox{ is differentiable} \right\} , \left( U,  \intle{n,k}\left( U\times \left( \RR ^k\right)  ^n\right)  \right), \mathrm{incl.}\right) \\ 
	&\in  &\SUBscone{\wCpo }{ \sconeFUNCTOR{n} } .
\end{eqnarray*} 

We define $\monadLR{n}{-}$ on $\SUBscone{\wCpo }{\sconeFUNCTOR{n}}$ by \eqref{eq:the-monad-on-subscone}  where $\unlift{\monadLR{n}{D, \left( C, C'\right) , j}} $ is the union \eqref{eq:the-underlying-definition-of-the-monad} with the full $\wCpo$-substructure of 
$ \sconeFUNCTOR{n}  \left( \monadwP{C} , \monadwP{C'} \right)  $ 
induced by the inclusion $\morLRmonad{X}	$ which is defined by the components given in \eqref{eq:monad-com1}, \eqref{eq:monad-com2}, and \eqref{eq:monad-com3}.
\normalsize
\begin{eqnarray}
	&&\left( \alpha _0,  \alpha_1 = \left( \beta _0 : U\to C, \beta _1 : \intle{n, k}\left( U\times \left( \RR ^k \right)  ^n\right) \to C' \right)    \right) \nonumber \\ &&\mapsto    \left( \lift{\beta _0} : \RR ^n\to \monadwP{C} , \lift{\beta _1} :\left( \RR \times \RR ^k\right)  ^n\to \monadwP{C'}  \right)\label{eq:definition-equation-monad-alpha0alpha1}\
\end{eqnarray}
\normalsize 
\begin{equation}\label{eq:the-monad-on-subscone}
	\monadLR{n}{D, \left( C, C'\right), j} \defeqq  \left( \unlift{\monadLR{n}{D, \left( C, C'\right) , j}} , \left( \monadwP{C} , \monadwP{C'}\right) , \morLRmonad{X}	
	\right) 
\end{equation}
\begin{equation} \label{eq:the-underlying-definition-of-the-monad}
	\left\{  \leastelement  \right\}
	\sqcup 	
	\sconeFUNCTOR{n}  \left( C , C'\right) 
	\sqcup 
	\left( \coprod _ {U\in\topologyO{n} } 
	\ehom{\SUBscone{\wCpo }{\sconeFUNCTOR{n}}}{\openLift{U, n} }{\left( D, \left( C, C'\right) , j\right)}\right)
\end{equation} 
\small 
\begin{enumerate}[$\mathfrak{c}$.1] 
	\item \label{eq:monad-com1} the inclusion $\left\{ \leastelement \right\} \to\sconeFUNCTOR{n}  \left( \monadwP{C} , \monadwP{C'} \right)  $ of the least morphism $\leastelement : \left( \RR ^n , \left( \RR  \times \RR ^k\right) ^n \right)\to \left( \monadwP{C}, \monadwP{C'} \right) $ in $\ehom{\wCpo\times\wCpo}{\left( \RR ^n , \left( \RR \times \RR ^k \right) ^n  \right) }{\left( \monadwP{C} , \monadwP{C'} \right) }$;
	\item \label{eq:monad-com2} the inclusion of the total functions $\sconeFUNCTOR{n}\left( \ee _{C} , \ee _{C'} \right) : \sconeFUNCTOR{n}  \left( C , C' \right)\to \sconeFUNCTOR{n}  \left( \monadwP{C} , \monadwP{C'} \right) $;
	\item \label{eq:monad-com3} for each $U\in \topologyO{n}$, the injection $\displaystyle\ehom{\SUBscone{\wCpo }{\sconeFUNCTOR{n}}}{\openLift{U, n} }{\left( D, \left( C, C'\right) , j\right)}\to \sconeFUNCTOR{n}  \left( \monadwP{C} , \monadwP{C'} \right) $ defined by \eqref{eq:definition-equation-monad-alpha0alpha1} 
	where 
	$\lift{\beta _0} $ and  $\lift{\beta _1} $ are the respective corresponding canonical extensions.
\end{enumerate} 
\normalsize

By lifting the multiplication and unit of $\monadwP{-}$,
the definition above gives us a strong monad $\monadLR{n}{-}$  on $\SUBscone{\wCpo }{\sconeFUNCTOR{n} }$ that is a lifting 
of $\monadwP{-}$ along 	the forgetful $\wCpo$-functor $\forgetfulSub _{n} : \SUBscone{\wCpo }{\sconeFUNCTOR{n} }\to \wCpo\times \wCpo$. Moreover, it is clear that $\monadLR{n}{-}$ 
satisfies the conditions of \ref{subsect:LR-as-rCBV-model}. Therefore, by Theorem \ref{theo:subsconing-generic},	$\left( \SUBscone{\wCpo }{\sconeFUNCTOR{n} } , \monadLR{n}{-} \right)$  is an $rCBV$ $\wCpo $-pair, and $\forgetfulSub _{n}$ yields an $rCBV $ $\wCpo $-pair morphism.
We have now effectively given a logical relations reasoning principle for derivatives of partially defined functions.

\section{Logical relations for ``$\reals$'' -- kickstarting the LR proof} 
We can, now, establish the logical relations' assignment. By the observation on differentiability, it is clear that we want to assign $\reals$ to the object \eqref{assig-object-LR} in $\SUBscone{\wCpo }{\sconeFUNCTOR{n }}$. While for each 
primitive operation $\op\in \Op _ m$ of the syntax,\footnote{We consider, here, the constants and $\tSign$ as well.} we have that the pair $\left( \sem{op} , \semt{k}{\dDSemtra{k}{\sem{\op}  }}\right)  $ defines a morphism 
$\semLR{n}{\op } : \semLR{n}{\reals }^m \to  \monadLR{n}{\semLR{n}{\reals }}$  in $\SUBscone{\wCpo }{\sconeFUNCTOR{n }}$ by the chain-rule for derivatives. Analogously, we define compatible morphisms $\semLR{n}{\tSign }$ and $\semLR{n}{\cnst c}$
by the morphisms in $\SUBscone{\wCpo }{\sconeFUNCTOR{n }}$ defined resp. by the pairs $\left( \signR ,   \dDSemtra{k}{\signR } \right)$ and $\left( \semanc{c} , \dDSemtra{k}{\semanc{c} }  \right)$.
\footnotesize \begin{equation}\label{assig-object-LR}
	\semLR{n}{\reals }\defeqq   \left(\left\{ \left( f : \RR ^n\to \RR , \secondM{f} \right) : f\mbox{ is differentiable, } \secondM{f} =  \dDSemtotaltra{k}{f} \right\}, \left( \RR , \RR \times \RR ^k \right), \mathrm{incl.}   \right)  
\end{equation}
\normalsize 
By the universal property of the syntactic $rCBV$ model  $\left( \SynVr, \SynTr, \SynffixpointRecT   \right)$, there is only one $rCBV$ model morphism $	\semLR{n}{-} $ compatible with the assignment above and, moreover, 
we can conclude that \eqref{eq:basic-diagram-commutativity-logical-relations} commutes.
\small\begin{equation}\label{eq:basic-diagram-commutativity-logical-relations}
	\diag{basic-logicalrelations-recursive-types}
\end{equation} \normalsize

\subsection{AD correctness from logical relations}
The first observation is that, indeed, our definitions give the desired predicate. More precisely, if  \textit{$ \left( h, \dot{h}\right) \in\unlift{\monadLR{n}{ \coprod\limits _{j\in L} \semLR{n}{\reals  } ^{l_j} }} $,
	then $ h : \RR ^n\to\monadwP{\coprod\limits _{j\in L} \RR ^{l_j}}  $ is differentiable and $\dot{h} = \dDSemtra{k}{h} $.}
By the definition of differentiable morphisms between coproducts of cartesian spaces, we have:
\begin{theorem}\label{coro:fundamental-LR-conclusion-about-morphisms-subscone}
	If, for each $i\in \KI$, the morphism   $	\left(h, \dot{h} \right) $ in $\wCpo\times\wCpo $ defines the morphism \eqref{eq:morphism-CBV-wCPO-pair-Subscone-defined} in $\SUBscone{\wCpo }{\sconeFUNCTOR{{s_i}}}$, then $ h : \coprod\limits _{r\in \KI} \RR ^{s_r}\to\monadwP{\coprod\limits _{j\in L} \RR ^{l_j}}  $ is differentiable and $\dot{h} = \dDSemtra{k}{h} $.\vspace{-12pt}
\end{theorem}  
\noindent\begin{minipage}{.6\linewidth}
	\begin{equation} \label{eq:morphism-CBV-wCPO-pair-Subscone-defined}
		\mathtt{h} : \coprod _{r\in \KI } \semLR{{s_i}}{\reals } ^{s _r}  \to \monadLR{{s_i}}{ \coprod _{j\in L}\semLR{{s_i}}{\reals }^{l _j} }
	\end{equation} 
\end{minipage}%
\noindent\begin{minipage}{.4\linewidth}
	\begin{equation} \label{eq:LR-the-coprojection-LR}
		\left( h\circ \coproje{\RR^{s_i}} ,  \dot{h}\circ\coproje{\left( \RR\times\RR ^k\right) ^{s_i}} \right)
	\end{equation} 
\end{minipage}	\normalsize
\begin{proof}
	For each $i\in \KI$, we have that \eqref{eq:LR-the-coprojection-LR} defines a morphism $\mathtt{h}_i $  is a morphism from $\semLR{{s_i}}{\reals } ^{s _i}\to\monadLR{{s_i}}{ \coprod _{j\in L}\semLR{{s_i}}{\reals }^{l _j} }$ in $\SUBscone{\wCpo }{\sconeFUNCTOR{{s_i}}}$. This implies that \eqref{eq:LR-the-coprojection-LR} belongs to $\unlift{\monadLR{s_i}{ \coprod\limits _{j\in L} \semLR{s_i}{\reals  } ^{l_j} }}$. By the observed above, this shows that $h\circ \coproje{\RR^{s_i}} $
	is differentiable and $\dot{h}\circ\coproje{\left( \RR\times\RR ^k\right) ^{s_i}} $ is its derivative. Since this results holds for every $i\in \KI$, the proof is complete.
\end{proof}

The commutativity of \eqref{eq:basic-diagram-commutativity-logical-relations} implies that, for any morphism $t  $ of the syntax (the category $\SynV $),
we have that the pair $\left(\sem{t}, \semt{k}{\DSyn\left(t \right) }  \right) $ defines a morphism
in $\semLR{n}{ t }$ in $\SUBscone{\wCpo }{\sconeFUNCTOR{{n}}}$ for every $n\in\NN $.
Therefore, by Theorem \ref{coro:fundamental-LR-conclusion-about-morphisms-subscone}, we have:
\begin{theorem}\label{theo:maybe-main-theorem}
	Let $t : \ty{1} \to \ty{2} $ be a morphism of $\SynV $, i.e. a program in our source language. If there are families $\left( s_r\right) _{r\in \KI} $ and $\left( l_j\right)_{j\in L} $ such that \eqref{eq:image-of-data-types-hypothesis} 
	and \eqref{eq:image-of-data-types-hypothesis2}
	hold for any $n\in \NN$ (where $\cong$ is just an isomorphism induced by coprojections and projections), then 	 $\sem{t}$ is differentiable and $\semt{k}{\DSyn\left(t \right) } = \dDSemtra{k}{\sem{t}}$.
\end{theorem} 	
\noindent\begin{minipage}{.5\linewidth}
	\begin{equation} \label{eq:image-of-data-types-hypothesis} 
		\semLR{{n}}{\ty{1} }  \cong  \coprod _{r\in \KI } \semLR{{n}}{\reals } ^{s _r} 
	\end{equation} 
\end{minipage}%
\noindent\begin{minipage}{.5\linewidth}
	\begin{equation} \label{eq:image-of-data-types-hypothesis2}
		\semLR{{n}}{\ty{2} } \cong  \coprod _{j\in L}\semLR{{n}}{\reals }^{l _j} 
	\end{equation} 
\end{minipage}	\normalsize\\
Since $ \semLR{{n}}{-} $ is an $rCBV$ model morphism, the hypothesis of Theorem~\ref{theo:maybe-main-theorem}  holds for
any data types that do not involve function types (including types built using recursion)  by Theorem \ref{theo:image-of-recursive-types-syntactic} of Appendix \ref{app:image-of-recursive-types}.
Therefore:
\begin{theorem}\label{theo:main-theorem-section-proof-recursive}
	Assume that $\tangentreals $ implements the vector space $\RR ^k$, for some $k\in\NN\cup\left\{ \infty \right\}$. For any program $\var{1}:\ty{1}\vdash \trm{1}:\ty{2}$  where	$\ty{1},\ty{2}$ are data types not involving function types in their construction, we have that $\sem{\trm{1}} $ is differentiable and, moreover, $
	\semt{k}{\Dsyn{\trm{1} }} = \dDSemtra{k}{\sem{\trm{1} }}$
	provided that $\Dsynsymbol $ is sound for primitives.
\end{theorem} 	 

\section{Forward vs reverse mode AD -- choosing $k$}\label{sec:correctness}
We have so far been vague about how to choose $k$ and whether we are considering forward or reverse AD. It turns out that our abstract development is enough for both AD methods, by choosing $k=1$ for forward (no need for vectorised AD) and $k=\infty$ (AD with dynamically sized vectorized tangent types) for reverse AD.
(Here, we remember that a practical implementation of dual numbers reverse AD like that of \cite{smeding2022} would make use of a distributive law as a runtime optimisation to reach the correct computational complexity.)

\subsection{Correctness of the dual numbers forward AD \;($k=1$)}\label{sub:forward-mode-types-correctness}
We assume that $\tangentreals$ implements the vector space $\RR$.
It is straightforward to see that we get forward mode AD out of our macro $\Dsynsymbol $: namely, for a program $\var{1}:\ty{1} \vdash \trm{1}:\ty{2} $ (where $\ty{1} $ and $\ty{2}$ are data types) in the source language, we get a program $\var{1}:\Dsyn{\ty{1}} \vdash \Dsyn{\trm{1}}:\Dsyn{\ty{2}}  $ in the target language, which, by Theorem \ref{theo:main-theorem-section-proof-recursive}, satisfies the following properties: \textbf{(A)} $\sem{\trm{1} } :  \coprod_{r\in K } \RR ^{n_r} \to\monadwP{ \coprod_{j\in L } \RR ^{m_j}   }   $ is differentiable; \textbf{(B)} if $y\in\RR ^{n_i}\cap\sem{\trm{1} }^{-1}\left( \RR ^{m_j} \right) = W_ j $ for some $i\in K $ and $j\in L $,  we have that,  \eqref{eq:forward-mode-equation} holds, for any $w\in \RR ^{n_i} $, where $\sem{\trm{1} }'(y) : \RR ^{n_i}\to\RR ^{m_j} $ is the derivative of $\sem{\trm{1} }|_{W_j}: W_j \to \RR ^{m_j} $ at $y$.
\begin{equation}\label{eq:forward-mode-equation}
	\semt{1}{ \Dsyn{\trm{1}} } \left( \intle{{n_i},1}\left(y,w\right) \right) = \intle{l,1}\left( \sem{\trm{1} }\left( y\right) ,  \sem{\trm{1} }'(y)(w) \right) 
\end{equation}

\subsection{Correctness of the dual numbers reverse AD \;($k=\infty$)}\label{sub:reverse-mode-types-correctness}
The following shows how our macro encompasses reverse mode AD.  We assume that $\tangentreals $ implements the vector space $\RR ^\infty $ (representing the case of a type of dynamically sized array of cotangents). 

For each $s\in\NN\cup\left\{\infty \right\} $, we consider the respective (co)projections 
$\semanticshandler{\infty}{s} $, and we define the morphism $\wrapSyncat{s}\defeqq \pairL \proj{j}, \cncanoni{j} \pairR _{j\in \NN } : \reals ^s\to \left( \reals\times \tangentreals \right) ^s $
in $\SynVt$.
For a program $\var{1}:\reals^{s} \vdash \trm{1}:\reals ^l$ (where $s, l\in \NN^\ast  $), we have that, for any $y\in\sem{\trm{1} }^{-1}\left( \RR ^l \right)\subset \RR ^s  $, \eqref{eq:transpose-derivative} holds
by Theorem \ref{theo:main-theorem-section-proof-recursive}. This gives the transpose derivative $\semanticshandler{s}{\infty } \sem{\trm{1} }'(y) ^t$ as something of the type $\tangentreals ^l $. 
The type can be fixed by using the handler $\tangentprojection{s}$ (see \cite{LUCATELLI-VAKAR-2022-Extended}).
\begin{equation}\label{eq:transpose-derivative}
	\semt{\infty }{\Dsyn{\trm{1}}\circ \wrapSyncat{s}  } \left( y \right) = \intle{l,\infty }\left( \sem{\trm{1} }\left( y\right) ,  \semanticshandler{s}{\infty } \sem{\trm{1} }'(y) ^t  \right)
\end{equation}	 
By Theorem \ref{theo:main-theorem-section-proof-recursive}, it is straightforward to generalize the correctness statements above to more general data types $\ty{2} $.

\section{Final remarks}
This work improved on the proof previously given in \cite{vakar2020denotational}:
we gave a simple correctness proof of dual numbers forward and reverse AD for realistic ML-family languages by making use of nimble new logical relations techniques for recursive types and partial differentiable functions.
In particular, we have simplified the argument to no longer depend on diffeological or sheaf-structure and to have it apply to arbitrary differentiable (rather than merely smooth) operations.
We have further simplified the subsconing technique for recursive types.

Although we can formulate the subsconing technique presented here in more general settings, such as the  setting of bilimit expansions \cite{levy2012call}, we opted for a simpler presentation -- making use of the $rCBV$ $\wCpo$-pairs introduced herein.
This approach is enough for semantic (open) logical relations, since we usually can interpret $CBV$ languages with recursive types in a simple enough  $rCBV$ $\wCpo$-pair. 
We believe that working with this special case of the semantics significantly simplifies our presentation.
We leave the 
presentation of the results in the setting of bilimit expansions for future work (if we find a useful setting where our current approach does not apply).

The use of subscone instead of the scone was a matter of presentation as well. Everything we did could be done for the scone (the comma category).
Although it takes more work to establish it, the subscone provides us with simpler verifications. It also gives the proof-irrelevant approach.

Finally, we hope that our work adds to the existing body of programming languages literature on automatic differentiation and recursion (and recursive types).
In particular, we believe that it provides a simple, principled denotational explaination of how AD and expressive partial language features should interact.
We plan to use it to generalise and prove correct the more advanced AD technique CHAD \cite{vakar2021chad,DBLP:journals/toplas/VakarS22,VAKAR-LUCATELLI2021} when applied to languages with partial features.
\section{Related Work}\label{sec:related-work}
There has recently been a flurry of work studying AD from a programming language point of view, a lot of it focussing on functional formulations of AD and their correctness. 
Examples of such papers are 
\cite{pearlmutter2008reverse,elliott2018simple,shaikhha2019efficient,brunel2019backpropagation,abadi-plotkin2020,DBLP:conf/fossacs/HuotSV20,DBLP:journals/pacmpl/MazzaP21,vakar2021chad,VAKAR-LUCATELLI2021,DBLP:journals/corr/abs-2101-06757,DBLP:journals/toplas/VakarS22,DBLP:journals/pacmpl/KrawiecJKEEF22,smeding2022}.
Of these papers, \cite{pearlmutter2008reverse,abadi-plotkin2020,DBLP:journals/pacmpl/MazzaP21,smeding2022} are particularly relevant as they also consider automatic differentiation of languages with partial features.
Here, \cite{pearlmutter2008reverse} considers an implementation that differentiates recursive programs and the implementation of \cite{smeding2022} even differentiates code that uses recursive types. 
They do not give correctness proofs, however.

The present paper can be seen as giving a correctness proof of the techniques implemented by \cite{smeding2022}.
\cite{abadi-plotkin2020} does give a denotational correctness proof of AD on a first-order functional language with (first-order) recursion.
The first-orderness of the language allows the proof to proceed by plain induction rather than needing logical technique.
\cite{DBLP:journals/pacmpl/MazzaP21} proves the correctness of basically the same AD algorithms that we consider in this paper when restricted to PCF with a base type of real numbers and a real conditional.
Their proof relies on operational semantic techniques.
Our contribution is to give an alternative denotational argument, which we believe is simple and systematic, and to extend it to apply to languages which, additionally, have the complex features of recursively defined datastructures that we find in realistic ML-family languages.

	%
	%
	%
	\bibliographystyle{splncs04}
	\bibliography{bibliography}

	\newpage
	\appendix
	
	\section{Source language}\label{app:Typing-Rules}
	\begin{figure}[!ht]
	\fbox{\parbox{0.98\linewidth}{\begin{minipage}{\linewidth}\noindent\input{type-system-shortversion}\end{minipage}}}
	\caption{Typing rules for the source language, where $\RRsyntax\subset\RR$ is a fixed set of real numbers containing $0$.
	Types can use type variables $\alpha,\beta$ from the kinding context $\Delta$.
		\label{fig:typestermrecursion-iteration}}
\end{figure}

\begin{figure}[!ht]
 \fbox{\parbox{0.98\linewidth}{\scalebox{0.92}{\begin{minipage}{\linewidth}\noindent\input{beta-eta1shortversion}\end{minipage}}}
}\caption{\label{fig:beta-eta} The standard $\beta\eta$-equational theory for a $CBV$ language with recursive types.
	We write $\freeeq{x_1,...,x_n}$ to indicate that the variables $x_1,\ldots,x_n$ are fresh in the
	left hand side.
	In the second rule, $x$ may not be free in $r$. Equations hold on pairs of terms of the same
	type.
}
\end{figure}
	\newpage\section{Typing rules for the target language}\label{app:Target-Typing-Rules}
	\begin{figure}[!ht]
	\fbox{\parbox{0.98\linewidth}{\begin{minipage}{\linewidth}\noindent\input{type-system-shortversion}\end{minipage}}}
	\caption{Typing rules for the source language, where $\RRsyntax\subset\RR$ is a fixed set of real numbers containing $0$.
		\label{fig:typestermrecursion-iteration}}
\end{figure}

\begin{figure}[!ht]

	\fbox{\parbox{0.98\linewidth}{\begin{minipage}{\linewidth}\noindent\input{type-system-target}\end{minipage}}}
	\caption{Extra typing rules for the target language, where we denote $\NN ^\ast := \NN - \left\{ 0 \right\}$, $\reals ^1 := \reals $ and $\reals ^{i+1} = \reals ^i \times \reals $.
		\label{fig:types-target-cotangent} }
\end{figure}

\begin{figure}[!ht]
	\fbox{\parbox{0.98\linewidth}{\scalebox{0.92}{\begin{minipage}{\linewidth}\noindent\input{beta-eta1shortversion}\end{minipage}}}
	}\caption{\label{fig:beta-eta} The standard $\beta\eta$-equational theory for a $CBV$ language with recursive types.
		We write $\freeeq{x_1,...,x_n}$ to indicate that the variables $x_1,\ldots,x_n$ are fresh in the
		left hand side.
		In the second rule, $x$ may not be free in $r$. Equations hold on pairs of terms of the same
		type.
	}
\end{figure}

\newpage\section{Image of recursive types}\label{app:image-of-recursive-types}
	While the logical relations for the primitive types are the primitive ones (defined by the image of each primitive object by $\SEMLRR{ - }{\ty{1} } $), we get the logical relations of more general data types out of the fact that $\SEMLRR{ - }{\ty{1} } $ is structure preserving. 
We finish this section establishing the result that underlies
the computation of the logical relations for data types in our setting.

We start by giving the definition corresponding to data/positive types for any $rCBV$ model with a chosen \textit{finite} set $\PTYPE$ of objects, playing the role of the set of primitive types. These are the objects
inductively defined by finite products, finite coproducts and recursion of objects in $\PTYPE$.
\begin{definition} 
	Let $ \left(\catV , \monadT, \ffixpointRecT \right) $ be an $rCBV$ model, and  $\PTYPE$ a finite set of objects of $\catV$.
	For each $K\in\PTYPE$, let $\pEE{\underDot{K}}, \pEE{I}, \pEE{O} : \catV ^\op\times \catV \to\catV $ be the constant functors which are, respectively, equal to $\underDot{K} $, $\terminall$ and $\initiall$.
	\textit{We define the set $\ParamVdtype{\catV , \monadT, \ffixpointRecT  }{\PTYPE } $ inductively by \ref{data-type-first}, \ref{data-type-second} and \ref{data-type-third}. }
	\begin{enumerate}[(D1)]
		\item \label{data-type-first} The functors $\underDot{K} , \pEE{I}, \pEE{O}$ are in $\ParamVdtype{\catV , \monadT, \ffixpointRecT  }{\PTYPE } $. Moreover, the projection $\proj{2}:\catV  ^\op\times \catV \to \catV $ belongs to $\ParamVdtype{\catV , \monadT, \ffixpointRecT  }{\PTYPE } $.
		\item \label{data-type-second} For each $n\in \NN ^\ast $, if the functors \eqref{eq:basic-functor-for-assumption-inductive-definition-data-types} belong to $\ParamVdtype{\catV , \monadT, \ffixpointRecT  }{\PTYPE } $, then the functors \eqref{eq:composition-of-Ev-diagonal} and \eqref{eq:step-doproduct-product-inductive}  are in $\ParamVdtype{\catV , \monadT, \ffixpointRecT  }{\PTYPE } $.
		\item \label{data-type-third} If $\pEE{E} = \left(\pE{E}{\catV}, \pE{E}{\catC} \right)\in\ParamVdtype{\catV , \monadT, \ffixpointRecT  }{\PTYPE } $ is such that $\pE{E}{\catV}\in\ParamVdtype{\catV , \monadT, \ffixpointRecT  }{\PTYPE } $, then 
		$ \left( \pE{\fixpointRecT{E}}{\catV } \right)$ is in
		$\ParamVdtype{\catV , \monadT, \ffixpointRecT  }{\PTYPE } $. 
	\end{enumerate}	
	\textit{We define the set  $\Positive{\catV , \monadT, \ffixpointRecT }{\PTYPE} $ of \emph{positive types} over $\PTYPE$ by \eqref{eq:definition-of-parametric-data-types}, that is to say, the subset of the $0$-variable parametric types in $\ParamVdtype{\catV , \monadT, \ffixpointRecT  }{\PTYPE }$.}
\end{definition} 
\noindent
\footnotesize \begin{minipage}{.6\linewidth}	
	\begin{equation}\label{eq:step-doproduct-product-inductive}
		\times \circ \left( G\times G'\right),  \sqcup \circ \left( G\times G'\right) : \left(\catV  ^\op\times \catV \right) ^{2n}  \to\catV
	\end{equation}
\end{minipage}\noindent
\begin{minipage}{.4\linewidth}	
	\begin{equation}\label{eq:composition-of-Ev-diagonal}
		\pEE{G}\circ \diagk{n} : \catV  ^\op\times  \catV \to\catV
	\end{equation} 	
\end{minipage}\\
\footnotesize\begin{minipage}{.5\linewidth}	
	\begin{equation}
		\pEE{G}, \pEE{G'} : \left(\catV ^\op\times \catV\right) ^n  \to\catV\label{eq:basic-functor-for-assumption-inductive-definition-data-types}
	\end{equation} 
\end{minipage}\noindent
\begin{minipage}{.5\linewidth}	
	\begin{equation}\label{eq:definition-of-parametric-data-types}
		\left\{ A \in \catV  : A\in\ParamVdtype{\catV , \monadT, \ffixpointRecT  }{\PTYPE }   \right\}
	\end{equation}
\end{minipage}\\

\normalsize
In the case of syntactic $rCBV$ models (see \ref{subsect:awesome-subsection-about-freely-fenerated-categ-structures}), taking $\PTYPE$ to be the set of primitive types, the definition above provides us with a formal definition of data types in our context.
By the universal property of these syntactic $rCBV$ models, we can establish the semantics and the logical relations as 
$rCBV$ model morphisms (as we do in \ref{subsect:semantics-source-language} and \ref{sec:logical-relations}). Hence, it is particularly useful to understand the image of (recursive) data types by $rCBV$ model morphisms.
The result below shows that the image of such a data type is always the coproduct of finite products of the image of the primitive types by the $rCBV$ model.

\begin{theorem}\label{theo:image-of-recursive-types}
	Let $\left(\catV , \monadT , \ffixpointRecT  \right) $ be an $rCBV$ model, and $\PTYPE$ a finite set of objects  of $\catV $. For each $D\in \Positive{\catV , \monadT, \ffixpointRecT }{\PTYPE} $, there is a countable family 
	$\displaystyle\left( \mind{j, K}\in\NN \right) _{(j, K)\in L\times\PTYPE } $ such that, for any $rCBV$ model morphism $H : \left( \catV, \monadT \right) \to \rCBVU\left( \catV ', \monadT ' \right)$, 
	$$ H(D)	\cong \coprod\limits _{j\in L}\left( \prod\limits _{K\in\PTYPE} H\left( K \right) ^\mind{j, K}  \right) , $$
	where the isomorphism $\cong $ is induced by coprojections and projections.\footnote{$\cong$ is induced by the universal property -- in other words, it is just a reorganization of the involved coproducts and products.} 
\end{theorem} 

Let $\left( \SynVr, \SynTr, \SynffixpointRecT   \right)$ be the syntactic $rCBV$ model established in Section \ref{sect:languages-as-rCBV-models}.
The positive types over $\reals$ are precisely those types corresponding to data types in our source language. Therefore:

\begin{theorem}\label{theo:image-of-recursive-types-syntactic}
	Let $\left(\catV , \monadT  \right) $ be an $rCBV$ $\wCpo$-pair. For each $\ty{1}\in\SynVr  $ corresponding to a (possibly recursive) data type, there is a countable family 
	$\displaystyle\left( s_r\in\NN \right) _{ r\in \KI } $ such that there is an isomorphism 
	$$ H(\ty{1} )	\cong \coprod\limits _{r\in s_r} H\left( \reals \right) ^{s_r}  , $$
	induced by coprojections and projections, provided that $H : \left( \SynVr, \SynTr, \SynffixpointRecT   \right)  \to \rCBVU\left( \catV , \monadT \right)$ is an $rCBV$ model morphism and $\left( \catV , \monadT \right)$ 
	is an $rCBV$ $\wCpo$-pair.
\end{theorem}

	
\end{document}